\documentclass[12pt,letterpaper]{article}
\usepackage[includeheadfoot, 
            marginratio={1:1,2:3}, 
            width=412pt, 
            height=688pt,]{geometry}
\usepackage{amsmath}
\usepackage{amsfonts}
\usepackage{amssymb}
\usepackage{bbm}

\usepackage{graphicx}

\allowdisplaybreaks[2]
\numberwithin{equation}{section}

\newcommand{\eq}[1]{\begin{equation}
                     \begin{split} #1 \end{split}
                     \end{equation}}
\newcommand{\eqa}[2]{\begin{equation}
                     \begin{array}{#1} #2 \end{array}
                     \end{equation}}
\newcommand{\ov}{\overline}

\newcommand{\thba}[2]{\bigl[ \begin{array}{@{}c@{}} \\[-6.5mm]
  {\scriptstyle #1} \\[-2.5mm]
  {\scriptstyle #2} \end{array}\bigr]}
\newcommand{\be}{\begin{equation}}
\newcommand{\ee}{\end{equation}}
\newcommand{\bea}{\begin{eqnarray}}
\newcommand{\eea}{\end{eqnarray}}
\newcommand{\mbb}{\mathbb}

\def\eins{\mathbbm{1}}
\def\ov#1{\overline{#1}}

\begin{document}

%%%%%%%%%%%%%%%%%%%%%%%%%%%%%%%%%%%%%%%%%%%%%%%
%%%%%%%%%%%%%%%%%%%%%%%%%%%%%%%%%%%%%%%%%%%%%%%
%%%%%%%%%%%%%%%%%%%%%%%%%%%%%%%%%%%%%%%%%%%%%%%
%%%%%%%%%%%%%%%%%%%%%%%%%%%%%%%%%%%%%%%%%%%%%%%
%%%%%%%%%%%%%%%%%%%%%%%%%%%%%%%%%%%%%%%%%%%%%%%
%%%%%%%%%%%%%%%%%%%%%%%%%%%%%%%%%%%%%%%%%%%%%%%
%%%%%%%%%%%%%%%%%%%%%%%%%%%%%%%%%%%%%%%%%%%%%%%
%%%%%%%%%%%%%%%%%%%%%%%%%%%%%%%%%%%%%%%%%%%%%%%

\vspace*{-1.5cm}
\begin{flushright}
  {\small
  MPP-2006-228 \\
  LMU-ASC 83/06 \\
  }
\end{flushright}

\vspace{1.5cm}
\begin{center}
  {\LARGE
  Non-perturbative SQCD Superpotentials \\
  from String Instantons \\
  }
\end{center}

\vspace{0.25cm}
\begin{center}
  {\small
  Nikolas Akerblom$^1$, Ralph~Blumenhagen$^1$, Dieter L\"ust$^{1,2}$, \\
  Erik Plauschinn$^{1,2}$ and Maximilian Schmidt-Sommerfeld$^1$ \\
  }
\end{center}

\vspace{0.1cm}
\begin{center}
  \emph{$^{1}$
  Max-Planck-Institut f\"ur Physik, F\"ohringer Ring 6, \\
  80805 M\"unchen, Germany } \\
  \vspace{0.25cm}
  \emph{$^{2}$ Arnold-Sommerfeld-Center for Theoretical Physics, \\
  Department f\"ur Physik, Ludwig-Maximilians-Universit\"at  M\"unchen, \\
  Theresienstra\ss e 37, 80333 M\"unchen, Germany} \\
\end{center}

\vspace{-0.1cm}
\begin{center}
  \tt{
  akerblom, blumenha, luest, plausch, pumuckl@mppmu.mpg.de \\
  }
\end{center}

\vspace{1.5cm}
\begin{abstract}
\noindent The Affleck--Dine--Seiberg instanton generated 
superpotential for SQCD with $N_f=N_c-1$ flavours 
is explicitly derived from
a local model of engineered intersecting 
D6-branes with a single E2-instanton. 
This computation extends also  to symplectic gauge groups
with $N_f=N_c$ flavours.
\end{abstract}

\thispagestyle{empty}
\clearpage
\tableofcontents

%%%%%%%%%%%%%%%%%%%%%%%%%%%%%%%%%%%%%%%%%%%%%%%
%%%%%%%%%%%%%%%%%%%%%%%%%%%%%%%%%%%%%%%%%%%%%%%
%%%%%%%%%%%%%%%%%%%%%%%%%%%%%%%%%%%%%%%%%%%%%%%
%%%%%%%%%%%%%%%%%%%%%%%%%%%%%%%%%%%%%%%%%%%%%%%
%%%%%%%%%%%%%%%%%%%%%%%%%%%%%%%%%%%%%%%%%%%%%%%
%%%%%%%%%%%%%%%%%%%%%%%%%%%%%%%%%%%%%%%%%%%%%%%
%%%%%%%%%%%%%%%%%%%%%%%%%%%%%%%%%%%%%%%%%%%%%%%
%%%%%%%%%%%%%%%%%%%%%%%%%%%%%%%%%%%%%%%%%%%%%%%

\section{Introduction}

Field theory and string theory can both be successfully 
dealt with in the perturbative  regime, where for the latter
we essentially only have a perturbative definition of the
theory.
However, in particular the vacuum structure of both theories
depends on the non-perturbative dynamics. In a semi-classical
approach such non-perturbative effects are generated by 
fluctuations around stationary topologically non-trivial
field configurations, such as instantons.

In general, these instantons can generate all kinds of effects,
but for four-dimensional ${\cal N}=1$ supersymmetric
theories their contribution to the holomorphic superpotential
can be understood quite explicitly. 
This has nicely been demonstrated by the celebrated
determination of the dynamically generated superpotential
for $SU(N_c)$ SQCD with  $N_f$ flavours by Affleck, Dine and Seiberg (ADS) \cite{Affleck:1983mk}.
In their paper holomorphy and various local and global symmetries like
gauge-, flavour- and R-symmetry were invoked to argue
that the superpotential can only have a very peculiar form,
whose coefficient was then shown to be non-vanishing
for the special case of $SU(2).$\footnote{Cf. also the explicit determination of prefactors by Cordes \cite{Cordes:1985um}.}

As mentioned above, instanton effects are also very important for an understanding of
the vacuum structure (these days called the landscape)
of string compactifications. Even though we do not have 
a non-perturbative definition of the theory, by analogy to field
theory rules for dealing with such string instantons have been
elaborated on  \cite{Dine:1986zy,Dine:1987bq,Becker:1995kb,Witten:1996bn,Ganor:1996pe,Green:1997tv,Harvey:1999as,Witten:1999eg,Buchbinder:2002ic,Beasley:2003fx,Beasley:2005iu,Buchbinder:2006xh}. The field theory
techniques are described in \cite{Dorey:2002ik,Terning:2006bq} (and references therein).
These instantons are given by wrapped
string world-sheets and by wrapped Euclidean D-branes and, like in field
theory, their contributions to the space-time superpotential are  quite
restricted. These contributions can be computed in a semi-classical approach, i.e.
one involving only the tree level instanton action and a one-loop
determinant for the fluctuations around the instanton.
Recently, in the KKLT scenario \cite{Kachru:2003aw}, 
wrapped D3-brane instantons have also played a major role in eventually fixing
all moduli.  For concrete orientifold realisations of KKLT see 
\cite{Denef:2005mm,Lust:2005dy,Lust:2006zg}.

For type IIA orientifold models on Calabi--Yau spaces with intersecting
D6-branes 
(and  their T-dual cousins) the
contribution of wrapped Euclidean D2-branes, hereafter called E2-branes, to the superpotential has been determined in \cite{Blumenhagen:2006xt} (see also \cite{dieter,Ibanez:2006da,Buican:2006sn}).
Since both the D6-branes and the E2-instantons are described
by an open string theory, it was shown that the entire
instanton computation boils
down to the evaluation of disc and 1-loop string diagrams with
boundary (changing) operators inserted. 
Here both the D6-branes and the E2-instantons wrap compact three-cycles
of the Calabi--Yau manifold. Note that in \cite{Blumenhagen:2006xt} the
instanton calculus was developed for the case that the
instantons do not lie on top of any stack of D6-branes, implying 
that only charged fermionic zero modes appear.

The natural question now arises whether one can consistently
recover known field theory instanton effects like the
ADS superpotential from string theory. 
Recall  that ${\cal N}=1$ SQCD with gauge group $SU(N_c)$
and $N_f$ flavours, $\Phi_{f}$, $\widetilde\Phi_{f'}$ 
with $N_f=N_c-1$ does indeed have a non-perturbatively
generated superpotential. 
In \cite{Affleck:1983mk} the authors argued that in this case the superpotential
is generated by a single field theory instanton and 
using various symmetries they restricted the form of  the superpotential
to be 
\bea
   W=\frac{\Lambda^{3N_c-N_f}}{{\rm det}[M_{ff'}] }\;, \nonumber
\eea 
with the meson matrix of rank $N_f$ defined as   
$M_{ff'}= \Phi_f^c\, \widetilde{\Phi}_{c,f'}$.
The above  question
has been posed in \cite{Florea:2006si} for the case of the runaway quiver gauge theory
of \cite{Intriligator:2005aw}. There the gauge theory  was $SU(3)$ with two quarks and 
the stringy origin of the constraint
$N_f=N_c-1$ was not really  obvious.
The above  question also relates to recent attempts at
freezing string moduli via flux induced superpotentials,
often extended by known non-perturbative superpotentials, which were actually only derived
for supersymmetric Yang--Mills theories. For such a simple
addition of these terms to make any sense, one must ensure that
string theory indeed  does generate these field theory superpotentials
by itself. Of course, one expects that the field theory result
can only be obtained in certain limits of string theory,
so one must know under what circumstances one can trust 
them.

Since such effects
scale like $\exp(-1/g^2_{SU(N_c)})$ in field theory, it is clear
that the E2-brane has to lie on top of the $U(N_c)$ stack
of D6-branes. In this case additional bosonic zero modes
appear which have to be  integrated over.
An E2-instanton inside a D6-brane is very much like the
D3-D($-$1) system, which is known to give a stringy realization of the ADHM instanton moduli 
space in the field theory limit \cite{Billo:2002hm}.

In this paper we show that applying the general formalism of \cite{Blumenhagen:2006xt}
to a locally engineered system of colour and flavour D6-branes
with an E2-instanton inside the colour branes indeed gives
the ADS superpotential in the field theory limit. Let us emphasise
that this result provides an important check on the rules proposed in \cite{Blumenhagen:2006xt}.
We note  that we are not using any a priori assumptions  to fix the shape
of the superpotential, but carry out the non-trivial evaluation
of the zero mode integrals in full glory. To our knowledge, such 
a direct computation has so far only been carried out in \cite{Cordes:1985um},
though with a different instanton zero mode measure than 
the ADHM measure which arises naturally in string theory. 
As already used in \cite{Florea:2006si}, the D-brane realisation of both the 
gauge theory and the instanton directly leads to 
extended quiver diagrams containing both the D6-branes and
the E2-instantons. This allows one to diagrammatically encode both 
the massless spectrum including the instanton zero modes
and possible allowed disc correlators directly in the quiver diagrams.
We easily recover the $N_f=N_c-1$ constraint from such a counting
of fermionic zero modes.

We also apply our techniques to the case of $USp(2N_c)$ SQCD 
with $N_f$ flavours and again obtain the known
field theory result for the dynamically generated superpotential \cite{Intriligator:1995ne}. 
We comment also on the slightly more involved $SO(N_c)$ case \cite{Intriligator:1995id}, which, 
in principle, is also amenable to our methods.

This paper is organised as follows:
In section 2 we first briefly recall the main results of \cite{Blumenhagen:2006xt}, which 
provide a recipe of how to compute contributions to the
superpotential by E2-brane instantons in a conformal field
theory framework. In addition we  derive some new results on the 
generalisation
of this framework  to E2-instantons carrying 
extra bosonic zero modes.  Moreover, we make a concrete
proposal relating the one-loop determinants for the fluctuations
around the instanton to one-loop gauge threshold corrections.   
Then we specify which E2-instantons
can be related to  field theory gauge instantons in a certain
field theory limit of string theory. That means, we locally 
engineer a brane set-up which corresponds to SQCD. 
In section 3 we compute the relevant disc amplitudes of 
the various boundary changing operators involving  both the
matter and the instanton zero modes. 
In section 4 we explicitly carry out the instanton zero mode
integration, which turns out to be non-trivial, and
eventually in the field theory limit we 
indeed arrive at the ADS superpotential.
In section 5 this result is generalised to supersymmetric
$USp(2N_c)$ gauge theory with $N_f$ flavours and in section 6
we give our conclusions.

%%%%%%%%%%%%%%%%%%%%%%%%%%%%%%%%%%%%%%%%%%%%%%%
%%%%%%%%%%%%%%%%%%%%%%%%%%%%%%%%%%%%%%%%%%%%%%%
%%%%%%%%%%%%%%%%%%%%%%%%%%%%%%%%%%%%%%%%%%%%%%%
%%%%%%%%%%%%%%%%%%%%%%%%%%%%%%%%%%%%%%%%%%%%%%%
%%%%%%%%%%%%%%%%%%%%%%%%%%%%%%%%%%%%%%%%%%%%%%%
%%%%%%%%%%%%%%%%%%%%%%%%%%%%%%%%%%%%%%%%%%%%%%%
%%%%%%%%%%%%%%%%%%%%%%%%%%%%%%%%%%%%%%%%%%%%%%%
%%%%%%%%%%%%%%%%%%%%%%%%%%%%%%%%%%%%%%%%%%%%%%%

\section{E2-instantons in Type II D-brane models}
\label{seca}

Let us review and slightly extend the formalism developed
in \cite{Blumenhagen:2006xt} for the computation of instanton
contributions to the superpotential of Type IIA
orientifolds with intersecting D6-branes and
E2-instantons wrapping compact three-cycles of the internal Calabi--Yau geometry.
We could also work with the possible T-dual Type IIB configurations,
but in order to have a clear geometric picture of
the arising open string zero modes we choose the Type IIA
framework.
We first discuss a compact set-up and then, as usual
for string engineering of field theories, take a decoupling
limit to eventually arrive at local geometries
comprising all the information relevant for the field
theory. For the techniques of model building with intersecting brane worlds in general, see for instance \cite{Uranga:2003pz,Blumenhagen:2005mu,Blumenhagen:2006ci}.

%%%%%%%%%%%%%%%%%%%%%%%%%%%%%%%%%%%%%%%%%%%%%%%
%%%%%%%%%%%%%%%%%%%%%%%%%%%%%%%%%%%%%%%%%%%%%%%
%%%%%%%%%%%%%%%%%%%%%%%%%%%%%%%%%%%%%%%%%%%%%%%
%%%%%%%%%%%%%%%%%%%%%%%%%%%%%%%%%%%%%%%%%%%%%%%

\subsection{Instanton generated Type IIA superpotential}

Assume we have a Type IIA orientifold with O6-planes
and intersecting D6-branes preserving ${\cal N}=1$
supersymmetry in four dimensions, i.e. 
the D6-branes wrap special Lagrangian (sLag) three-cycles
$\Pi_a$ of the underlying Calabi--Yau manifold, all preserving
the same supersymmetry.

Space-time instantons are given also by D-branes, which
in this case are Euclidean E2-branes wrapping three-cycles $\Xi$ in the Calabi--Yau, so that they
are point-like in four-dimensional
Minkowski space. Such instantons can contribute
to the four-dimensional superpotential, if they
preserve half of the ${\cal N}=1$ supersymmetry,
which means they also wrap sLag three-cycles.
This guarantees that there are at least the four translational
zero modes $x_\mu$ and two fermionic zero modes $\theta_i$
arising from the two broken supersymmetries.
If the instanton is not invariant under the orientifold
projection there are another two fermionic zero modes $\ov \theta_i$, as the instanton breaks four
of the eight supercharges in the bulk \cite{IbanezUranga}.
We also require that there do not arise
any further zero modes from E2--E2 and E2--E2' open strings,
so that in particular the three-cycle $\Xi$ should be rigid, i.e. $b_1(\Xi)=0$.

Therefore, considering  an E2-instanton in an 
intersecting brane configuration, additional zero modes can only arise
from the intersection of the instanton $\Xi$ with D6-branes $\Pi_a$:
\begin{itemize}

\item
At any rate, there are  $N_a\, [\Xi\cap \Pi_a]^+$ chiral fermionic
zero modes $\lambda_{a,I}$ and $N_a\, [\Xi\cap \Pi_a]^-$
anti-chiral ones, $\widetilde{\lambda}_{a,J}$.\footnote{Here we introduced 
the physical intersection number between two branes $\Pi_a\cap\Pi_b$, which 
is the sum of positive $[\Pi_a\cap \Pi_b]^+$ and negative 
$[\Pi_a\cap \Pi_b]^-$ intersections.} 

\item
At a smooth point in the CY moduli space, 
in general there will be no bosonic zero modes originating form the instanton. However, if the 
instanton lies right on top of a D6-brane one gets $4\, N_a$ 
bosonic zero modes. Let us organize them into two complex modes $b_{\alpha}$ with $\alpha=1,2$. 

\end{itemize}

\noindent
As we will see, for $\Xi=\Pi_a$  
there can appear extra conditions
on the zero modes arising from the flatness
conditions for the effective zero-dimensional
gauge theory on the E2-instanton.
In this case the above numbers only give the
unrestricted number of zero modes.

In \cite{Billo:2002hm,Blumenhagen:2006xt} it was argued that each E2--D6 zero mode
carries an extra normalisation factor of $\sqrt{g_s}$
so that contributions to the superpotential
can only arise from CFT disc and one-loop
diagrams involving E2 as a boundary. Here
the disc contains precisely two
zero mode insertions  and the 1-loop diagram
none. For details we refer the reader
to \cite{Blumenhagen:2006xt}, here we only recall  the final
result.

For its presentation it is useful to  introduce the short-hand notation
\bea
  \widehat\Phi_{a_k,b_k}[{\vec x_k}] = \Phi_{a_k,x_{k,1}}\cdot 
  \Phi_{x_{k,1},x_{k,2}} \cdot \Phi_{x_{k,2},x_{k,3}}\cdot \ldots
  \cdot \Phi_{x_{k,n-1},x_{k,n}}  \cdot \Phi_{x_{k,n(k)},b_k}  
\eea
for the chain-product of open string vertex operators.
We define $\widehat\Phi_{a_k,b_k}[\vec 0]=\Phi_{a_k,b_k}$.
The single E2-instanton contribution
to the superpotential can be determined 
by evaluating the following zero mode integral over disc and 1-loop
open string CFT amplitudes
\eq{
  \label{general_w}
  S_W=&{1\over \ell_s^3}\, \int d^4 x\, d^2\theta \,\,
    \sum_{\rm conf.}\,\, 
    {\textstyle 
    \prod_{I}  d\lambda_I\, \prod_{J} d\widetilde{\lambda}_J  
    \prod_{\alpha=1}^2
    \prod_{i}  db_{\alpha,i}\, d\ov{b}_{\alpha,i}  } 
    \;
    \\
  & \quad\times
    \delta(\lambda,\widetilde{\lambda},b_{\alpha},\ov b_{\alpha} )
    \times\exp \bigl( \langle 1 \rangle^{\text{1-loop}} \bigr) 
    \times\exp \left( \langle 1 \rangle^{\rm disc} \right)
    \\ 
  & \quad\times
    \exp\left(\sum_k \, \langle \langle \Phi_{a_k,b_k}[\vec x_k]  
    \rangle\rangle^{\rm disc}_{\lambda_k\widetilde{\lambda}_k} \right)
    \times
    \exp\left(\sum_l\, \langle\langle \Phi_{a_l,b_l}[\vec x_k]  
    \rangle\rangle^{\rm disc}_{b_{\alpha,l}\ov b_{\alpha,l}} \right) 
    \;,
}  
where we work in the convention that all fields carry no scaling
dimension and the sum is over all possible vertex operator insertions.
Note that in this formula the $\ov \theta_i$ zero modes
have been integrated over leading to $\delta$-functions representing
the fermionic ADHM constraints \cite{Atiyah:1978ri}, as will
be explained later on. There are further $\delta$-functions involving only bosonic zero modes. 
These incorporate the bosonic ADHM constraints or, in more physical terms, the  F-term
and D-term constraints from the effective theory
on the E2-brane. 
Since we will not attach any charged matter fields
to the one-loop diagrams, compared to \cite{Blumenhagen:2006xt}, we left out
this term in (\ref{general_w}).
The prefactor $\ell_s^{-3}$ is chosen for dimensional reasons
and is so far only determined up to numerical factors. (We will come back to
such normalisation issues later.) 
The vacuum disc amplitude is given by
\bea
    \langle 1 \rangle^{\rm disc}=-S_{{\rm inst}}
    =-{1\over g_s}{V_{\rm inst}\over 
    \ell_s^3}=-{8\pi^2\over g_{\rm YM}^2}\;,
\eea
where the four-dimensional Yang--Mills gauge coupling is meant to be for the
gauge theory on a stack of  D6-branes
wrapping the same internal three-cycle as the instanton.
As already mentioned,  in \cite{Blumenhagen:2006xt} 
the superpotential was given for the case
without any additional charged bosonic  zero modes $b_\alpha$, $\ov b_\alpha$.
Since they also carry the extra normalisation factor $\sqrt{g_s}$, 
precisely two of them can be attached to a disc diagram. Moreover, since
they commute the whole exponential term appears in the
superpotential (\ref{general_w}).

The one-loop contributions are annulus diagrams for open strings
with one boundary on the E2-instanton and the other 
boundary on the various D6-branes
\bea
   \label{loop_diagram}
   \langle 1 \rangle^{\text{1-loop}}=  
   {\textstyle \sum_a \left[ {Z'}^A ({\rm E2},{\rm D6}_a)
   +{Z'}^A ({\rm E2},{\rm D6}'_a)\right] +  {Z'}^M({\rm E2},{\rm 
   O6})}\;. 
\eea
Here $Z'$ means that we only sum over the massive open
string states in the loop amplitude, as the zero modes
are taken care of explicitly. Without soaking up these
zero modes, one clearly would encounter
divergences from integrating out bosonic zero modes
and zeros from integrating out fermionic zero modes. 
We have provided the formula for an orientifold model on a compact Calabi--Yau
manifold.
For the local engineering of SQCD done later in this paper, 
we do not really have to take the orientifold planes and the image branes
into account, so that the  annulus 
amplitudes with the image branes and the M\"obius strip amplitudes are absent.

The last ingredients to compute the rhs of equation (\ref{general_w}) are the disc 
diagrams with insertion of matter fields $\Phi_{a,b}$ and fermionic or
bosonic zero modes. In the case of inserting fermionic zero modes $\lambda$ 
and $\widetilde{\lambda}$, the explicit form of the disc diagram is the following
\begin{multline}
  \label{disc_diagram}
  \langle\langle \Phi_{a_k,b_k}[\vec x_k] \rangle
  \rangle^{\rm disc}_{\lambda_k\widetilde\lambda_k} = \\  
  \int {dz_1\ldots dz_{n+2}\over V_{1,2,n+2}} \; 
  \langle V_{\lambda_k}(z_1)\, V_{\Phi_{a_k,x_{k_1}}}(z_2)\,
  \ldots V_{\Phi_{x_{k_n},b_{k}}}(z_{n+1})\, 
  V_{\widetilde\lambda_k}(z_{n+2}) \rangle \;.
\end{multline}
For the disc diagram with insertion of bosonic zero modes $b_{\alpha}$ and $\ov{b}_{\alpha}$, one simply replaces $V_{\lambda_k}(z_1)$ by $V_{b_{\alpha,k}}(z_1)$ and $V_{\widetilde\lambda_k}(z_{n+2})$ by $V_{\ov{b}_{\alpha,k}}(z_{n+2})$. In equation \eqref{disc_diagram} the symbol $V_{ijk}$ denotes the measure 
\bea
   V_{ijk}={dz_i\,dz_j\,dz_k\over (z_i-z_j)(z_i-z_k)(z_j-z_k)}
\eea
with $z_i\to \infty$, $z_j=1$ and $z_k=0$. The remaining $z$'s are to be
ordered as $1\ge z_3\ge z_4\ge\ldots \ge z_{n+1}\ge 0 $ and integrated over the interval $[0,1]$.

\bigskip
To summarise, the entire computation of the single instanton
contribution to the superpotential boils down
to the following steps:
\begin{itemize}

\item 
Determining the bosonic and fermionic instanton
zero modes from all possible open string sectors.

\item
Finding the E2-flatness contraints for instantons
sitting on top of D6-branes and implementing them as delta-functions in
\eqref{general_w}.

\item
The  evaluation of certain disc amplitudes with 
insertion of precisely two instanton zero modes and an arbitrary number of 
matter fields (see equation \eqref{disc_diagram}). 

\item
The computation of vacuum annulus and M\"obius strip
amplitudes as indicated in equation \eqref{loop_diagram}.

\item
Performing the integration over bosonic and fermionic
zero modes.

\end{itemize} 
As expected, the stringy superpotential is entirely given  
in a semi-classical approximation. We will see that in the field theory limit, only the leading order terms of the disc and one-loop amplitudes survive, considerably simplifying formula \eqref{general_w}.

%%%%%%%%%%%%%%%%%%%%%%%%%%%%%%%%%%%%%%%%%%%%%%%
%%%%%%%%%%%%%%%%%%%%%%%%%%%%%%%%%%%%%%%%%%%%%%%
%%%%%%%%%%%%%%%%%%%%%%%%%%%%%%%%%%%%%%%%%%%%%%%
%%%%%%%%%%%%%%%%%%%%%%%%%%%%%%%%%%%%%%%%%%%%%%%

\subsection{One-loop amplitudes}

So far we have not really specified what we actually mean
by the vacuum one-loop amplitudes, i.e. the annulus
amplitudes for open strings between the
E2-instanton and the various D6-branes.
To our knowledge such amplitudes have not been computed
yet. So far, only overlaps of two non-instantonic or
two instantonic \cite{Green:1996um,Green:1997tv,Gutperle:1997iy} D-branes have been considered
in the literature \footnote{Just when the present paper was readied for submission,
we received \cite{abel06}, where these E2-D6 partition
functions were computed for intersecting branes on $\mbb T^6$.}.

\begin{figure}[t]
\begin{center}
  \includegraphics[width=0.7\textwidth]{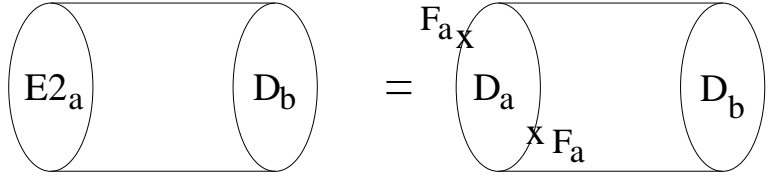}
\end{center}
\caption{Relation between instantonic one-loop amplitudes and 
corresponding
  gauge  threshold corrections.}\label{nicerel}
\end{figure}

In order to get an idea beforehand what they should be, we first note that
the divergence from the massless open string excitations
should be identified with  the  divergence from integrating
out naively the instanton zero modes, i.e. without soaking
them up in disc diagrams involving other space-time fields.
Next we observe
that for the case in which we place an ${\rm E2}_a$-instanton on top of a 
${\rm D6}_b$-brane,
the disc instanton action
\bea
     S_{\rm inst}={8\pi^2\over g_{\rm YM}^2}
\eea
is equal to the tree level gauge coupling on ${\rm D6}_b$.
This gauge coupling receives one-loop threshold
corrections from open strings between the brane
${\rm D6}_a$ and the branes ${\rm D6}_b$ of the form \cite{Lust:2003ky}
\bea
    {8\pi^2\over g^2_{a}(\mu)}={8\pi^2\over g^2_{a}(\mu_0)}
    + \sum_b b^a_{[ab]}\, \ln\left(\mu\over \mu_0\right) + 
    \Delta^a_{[ab]},
\eea
where $b^a_{[ab]}$ is the field theory one-loop beta function
coefficient and comes from massless states of the D6$_a$-D6$_b$ open
string sector running in the loop.
To make the picture consistent, it should better be that
the one-loop contributions $Z^A({\rm E2}_a,{\rm D6}_b)$ are identical
to the one-loop gauge threshold corrections $T^a({\rm D6}_a,{\rm D6}_b)$.
More concretely, the sum of the three even spin structures
will be shown to yield the one-loop correction to the gauge coupling, and the
odd spin structure is expected to give the one-loop correction to the $\theta$-angle.
We therefore propose that diagrammatically we have the intriguing
relation shown in figure \ref{nicerel}.
\noindent
To prove this statement let us first compute the instantonic
one-loop partition functions $Z^A({\rm E2}_a,{\rm D6}_b)$.
These cannot be expressed in light cone gauge, but including the 
contributions
from the ghosts can, for the even spin structures, be written as
\eq{
\label{instpart}
   Z^A({\rm E2}_a,{\rm D6}_b)=-\int_0^\infty  {dt\over t}\,
   \sum_{\alpha,\beta \neq (\frac{1}{2},\frac{1}{2})} 
   (-1)^{2(\alpha+\beta)}\,
    { \Theta^2\thba{\alpha}{\beta}(it,it/2) \over
      \Theta^2\thba{1/2}{1/2}(it,it/2)}\, {\eta^3(it) \over \Theta 
\thba{\alpha}{\beta}(it,0)}
    \,A^{\rm CY}_{ab}\thba{\alpha}{\beta}
}
with
\bea
    \Theta \thba{\alpha}{\beta}(\tau,z)=\sum_{n\in \mbb Z} e^{i\pi\tau 
(n+\alpha)^2}\, e^{2\pi i (n+\alpha)(z+\beta)} \; .
\eea
Moreover,  $A^{\rm CY}_{ab}\thba{\alpha}{\beta}$ denotes the internal open 
string
partition function with the respective spin-structure
for open strings between branes wrapping
the three-cycles $\Pi_a$ and $\Pi_b$.
Using the relations
\bea
             \left( \Theta \thba{\alpha}{\beta}(it,z)\, 
\Theta'\thba{1/2}{1/2}(it,0) \over
            \Theta \thba{1/2}{1/2}(it,z)\, 
\Theta\thba{\alpha}{\beta}(it,0)\right)^2=
           {\Theta''\thba{\alpha}{\beta}(it,0)\over \Theta 
\thba{\alpha}{\beta}(it,0)}-
             \partial^2_z \log \Theta\thba{1/2}{1/2}(it,z)
\eea
where the derivatives are taken with respect to the variable $z$, and $-2\pi 
\eta^3(it)=\Theta'\thba{1/2}{1/2}(it,0)$,
after a few standard manipulations we
can bring the partition function into the form
\bea
\label{partin2}
   Z^A({\rm E2}_a,{\rm D6}_b)=\int_0^\infty  {dt\over t}\,
   \sum_{\alpha,\beta\neq (\frac{1}{2},\frac{1}{2})} 
   (-1)^{2(\alpha+\beta)}\,
   {\Gamma \thba{\alpha}{\beta}(it) \over\eta^3(it)}\,\,
   A^{\rm CY}_{ab}\thba{\alpha}{\beta} \;.
\eea
The $\Gamma$'s are nothing else than the derivatives of
$\Theta$-functions with respect to the variable $t$
\bea
   \Gamma\thba{\alpha}{\beta} (it)=-{1\over \pi}\,
   \frac{\partial}{\partial t}
   \Theta\thba{\alpha}{\beta}(it,0)=\sum_{n\in \mbb Z}
   {\textstyle (n+\alpha)^2}\, e^{-\pi t (n+\alpha)^2}\,
   e^{2\pi i (n+\alpha) \beta} \;.
\eea

The threshold corrections for intersecting D6-branes on
a torus $\mbb T^6$ have been explicitly computed in \cite{Lust:2003ky}.
Their result is easily generalised to the case of intersecting
D6-branes on a general Calabi--Yau space and precisely gives (\ref{partin2}).

Now, divergences in  (\ref{partin2}) can arise from massless 
states.
As we mentioned, these divergences are precisely related
to the one-loop beta-function coefficient $b^a_{[ab]}$ for the
${\cal N}=1$ supersymmetric massless states between
the branes ${\rm D6}_a$ and ${\rm D6}_b$.
This one-loop divergence is roughly
\bea
       \exp\left( b^a_{[ab]}\, \int_{t_0}^t {dt\over t} q^0 \right)
         = \left(t\over t_0\right)^{b^a_{[ab]}}=
         \left(\mu_0\over \mu\right)^{b^a_{[ab]}},
\eea
where we have defined the mass scales $\mu=1/t$ and $\mu_0=1/t_0$.
The beta function coefficient $b^a_{[ab]}$ has contributions
from the NS- and the R-sector in (\ref{partin2})
\bea
        b^a_{[ab]}=\left( b^a_{[ab]}\right)_{NS}-
                 \left( b^a_{[ab]}\right)_{R}.
\eea
Identifying, as proposed, $T^a({\rm D6}_a,{\rm D6}_b)$
with
$Z^A({\rm E2}_a,{\rm D6}_b)$, these divergences \linebreak should come 
from integrating
out bosonic and fermionic instanton zero-modes.
Taking into account that a mass term in the action for a boson reads
${\cal L}^B_{\rm mass}=\mu^2 b\,\ov b $ and for a fermion
${\cal L}^F_{\rm mass}=\mu\, \lambda\, \widetilde\lambda$,
the number of bosonic and fermionic zero modes
should be identified with
\bea
       n_B=\left( b^a_{[ab]}\right)_{{\rm NS}}, \quad
       n_F=2\, \left( b^a_{[ab]}\right)_{{\rm R}}.
\eea
Let us verify  that this indeed gives the correct  number
of instanton zero modes for the configurations relevant
in the following:

\begin{itemize}

\item{\underbar{$a\ne b$}:
Let us discuss the case that we have an intersection giving rise
to one pair of chiral superfields between the ${\rm D6}_a$ and a stack
of $N_b$ ${\rm D6}_b$-branes.
In the NS sector, for the massless states
the internal sector carries $h=1/2$ and the
external one only the ground state with $h=0$.
However, in the threshold corrections this latter
$n=0$ state does not survive and therefore all states from
the NS-sector are massive and $(b_{[ab]}^a)_{{\rm NS}}=0$.
In the R-sector with $\alpha=1/2$, the $(n+\alpha)^2$ factor
implies that we get  $(b_{[ab]}^a)_{{\rm R}}=N_b$, leading to
  the correct beta-function coefficient for $N_b$ generations
of quarks. The number of fermionic zero modes for  the
${\rm E2}_a$ instanton and the $N_b$ D6-branes is therefore
\bea
\label{zeronn}
       n_B=0\,, \quad\quad n_F=2\, N_b\,.
\eea
}

\item{\underbar{$a=b$}:
Now let us discuss the case that we put the instanton
${\rm E2}_a$ on top of the stack of ${\rm D6}_a$-branes.
If the ${\rm D6}_a$ brane is rigid, we get precisely one
${\cal N}=1$ vector superfield in the corresponding
${\rm D6}_a$-${\rm D6}_a$ open string sector.
These are the states with $n=\pm 1$ in the external sector.
Therefore, in this case we find $(b_{[aa]}^a)_{{\rm NS}}=4\,N_a$
in (\ref{partin2}). The fermionic contribution comes
out as $(b_{[aa]}^a)_{{\rm R}}=N_a$, giving the correct
beta-function coefficient for one ${\cal N}=1$
vector field $b_{[aa]}^a=3\, N_a$.
The number of instanton zero modes therefore is
as expected
\bea
\label{zeromm}
       n_B=4\, N_a\, , \quad\quad n_F=2\, N_a\,.
\eea
}
\end{itemize}

We see that in these two examples indeed 
the E2--D6 partition functions are identical to the gauge threshold
correction in the corresponding D6--D6 sector.\footnote{Note that the superpotential in the Wilsonian sense is a holomorphic quantity but the stringy gauge threshold corrections in eq. \eqref{partin2} generically are not. Therefore, only the Wilsonian holomorphic part of the threshold corrections enters the instanton generated superpotential.}

%%%%%%%%%%%%%%%%%%%%%%%%%%%%%%%%%%%%%%%%%%%%%%%
%%%%%%%%%%%%%%%%%%%%%%%%%%%%%%%%%%%%%%%%%%%%%%%
%%%%%%%%%%%%%%%%%%%%%%%%%%%%%%%%%%%%%%%%%%%%%%%
%%%%%%%%%%%%%%%%%%%%%%%%%%%%%%%%%%%%%%%%%%%%%%%

\subsection{Engineering of field theory instantons}

The formalism presented so far deals with truly  
string instantons for compact Type IIA
orientifolds. In the following we would like
to extract the instanton amplitudes for
field theory instantons in supersymmetric Yang--Mills
theories.
This means that we have to decouple gravity and
higher order string corrections.
Field theory instantons in a supersymmetric 
Yang--Mills theory contain a factor $\exp(-1/g^2_{\rm YM})$, 
which immediately implies that such effects
can only arise
from E2-branes sitting right on top of the D6-brane
supporting the Yang--Mills theory.

To be concrete, let us engineer, in the spirit of \cite{Katz:1996fh,Katz:1996th}, $SU(N_c)$ supersymmetric 
Yang--Mills theory with $N_f$ non-chiral flavours.
This theory is generally called SQCD.
First, we take a stack of $N_c$ D6-branes wrapping a three-cycle $\Pi_c$
of some Calabi--Yau manifold. In order to just get ${\cal N}=1$ super
Yang--Mills theory on this brane, we assume that the three-cycle
is special Lagrangian and rigid, i.e. that the number
of its deformations satisfies $b_1(\Pi_c)=0$.
In D-brane models, fundamental representations of matter fields
can only come from bi-fundamental representations arising from
the intersection with  a stack of
$N_f$ flavour branes.  These flavour branes wrap a different
sLag three-cycle $\Pi_f$. 
As we will discuss in subsection \ref{secfield}, we would like the 
$U(N_f)$ gauge dynamics on these branes to decouple.

Since we want non-chiral matter\footnote{In toroidal $\mathbb{T}^2\times\mathbb{T}^2\times\mathbb{T}^2$ compactifications with intersecting branes this can be achieved for instance by choosing intersection angles $\phi_1=0$ and $\phi_2=-\phi_3$ between $\Pi_c$ and $\Pi_f$.}, we assume that the intersection number
of $\Pi_c$ and $\Pi_f$ is non-chiral, which means
\bea
         [\Pi_c\cap \Pi_f]^\pm=1 \;.
\eea
With this particle content, all cubic gauge anomalies
vanish so that we can assume that these two D-branes do not have any
other non-trivial intersection with the other potentially present
D6-branes in the model. Generically
the gauge boson in the $U(1)\subset U(N_c)$  subgroup acquires a mass due 
to GS mixing terms with the axions.

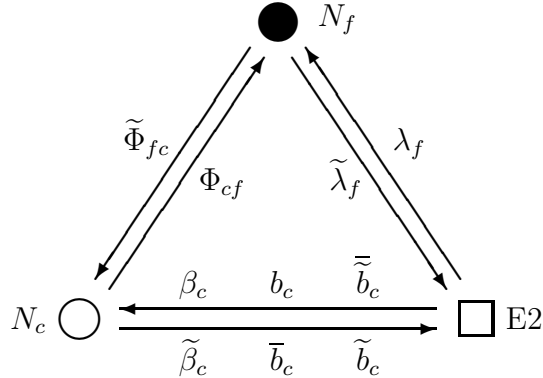
\begin{figure}[ht]
\begin{center}

\setlength{\unitlength}{0.75pt}
\begin{picture}(270,200)(-135,-25)

\thicklines\put(-100,0){\circle{20}}
\thicklines\put(0,150){\circle*{20}}
\thicklines\put(92,-8){\framebox(16,16){}}

\thicklines\put(-80,-5){\vector(1,0){160}}
\thicklines\put(80,5){\vector(-1,0){160}}
\thicklines\put(-85,15){\vector(2,3){78}}
\thicklines\put(-14,137){\vector(-2,-3){78}}
\thicklines\put(7,132){\vector(2,-3){78}}
\thicklines\put(92,20){\vector(-2,3){78}}

\put(-135,-5){$N_c$}
\put(115,-5){$\mbox{E2}$}
\put(20,148){$N_f$}

\put(-50,-26){$\widetilde{\beta}_c\qquad\overline{b}_c\qquad\widetilde{b}_c$}
\put(-50,13){$\beta_c\qquad b_c\qquad\overline{\widetilde{b}}_c$}

\put(-78,85){$\widetilde{\Phi}_{fc}$}
\put(-40,65){$\Phi_{cf}$}

\put(26,65){$\widetilde{\lambda}_f$}
\put(58,85){$\lambda_f$}

\end{picture}

\caption{Local SQCD brane configuration with resulting zero modes.\label{figa}}
\end{center}
\end{figure}

As we mentioned above, in string theory, 
the (zero size) Yang--Mills instanton in $SU(N_c)$ is given
by a Euclidean E2-brane wrapping the same three-cycle $\Pi_c$ as the
colour branes
and being point-like in four-dimensional Minkowski space.
The final local D6--E2 brane configuration and the 
resulting zero modes are shown 
in the extended quiver diagram in figure \ref{figa}.

\noindent
In this  local model, there are three open string sectors involving the 
instanton. We analyse them in turn.

%%%%%%%%%%%%%%%%%%%%%%%%%%%%%%%%%%%%%%%%%%%%%%%
%%%%%%%%%%%%%%%%%%%%%%%%%%%%%%%%%%%%%%%%%%%%%%%

\paragraph{E2--E2:} 
Since the E2-instanton wraps the
same rigid three-cycle as the D6$_c$ brane, the massless E2--E2 modes are 
the four positions in Minkowski space $x_\mu$ with $\mu=0,\ldots,3$ and 
four massless fermions $\theta_i$, $\ov\theta_i$, $i=1,2$.
The latter correspond to the breaking of four of the eight
supercharges preserved in type IIA theory on a Calabi--Yau manifold. Two of these,
namely the $\theta_i$, are related to the two supersymmetries preserved by the
D6-branes, but broken by the instanton, and, together with the four bosons $x_\mu$
contribute the measure
\bea
      \int d^4x\, d^2\theta
\eea
to the instanton amplitude. Integrating the other two zero modes
$\ov\theta_i$ leads to the fermionic ADHM \cite{Atiyah:1978ri} constraints
\cite{Billo:2002hm,Argurio:2007vq} as we will explain in the following.

%%%%%%%%%%%%%%%%%%%%%%%%%%%%%%%%%%%%%%%%%%%%%%%
%%%%%%%%%%%%%%%%%%%%%%%%%%%%%%%%%%%%%%%%%%%%%%%

\paragraph{E2--D6$_c$:} Since the E2-instanton and the ${\rm D6}_c$ branes wrap the same three-cycle, they share three directions in the internal space and have N-D boundary conditions along Minkowski space. Therefore, the ground state energy in both the NS- and the R-sector is $E_{\rm NS,R}=0$ and one gets $N_c$ complete hypermultiplets of massless modes. These are the $4\, N_c$ bosonic zero modes already seen in eq. (\ref{zeromm}).

However, not all of these bosonic zero modes are independent. In fact as analysed in detail in \cite{Billo:2002hm}, the effective zero-dimensional supersymmetric gauge theory on the E2-brane gives rise to D-term and F-term constraints, which in the field theory limit are nothing else than the well known ADHM \cite{Atiyah:1978ri} constraints \cite{Dorey:2000ww}.
If we rename the complex bosonic zero modes $b_{\alpha,i}$ as $b_{1,i}=b_i$ and $b_{2,i}=\ov{\widetilde{b}}_i$, these constraints become \cite{Khoze:1998gy}
\eq{
  \label{adhm_bos}
  \sum_{c=1}^{N_c} 
  \left[b_c\,\widetilde{b}_c+\ov{b}_c\,\ov{\widetilde{b}}_c
    \right]=0\;,
  \quad 
  \sum_{c=1}^{N_c} 
  \left[b_c\,\widetilde{b}_c-\ov{b}_c\,\ov{\widetilde{b}}_c
    \right]=0\;,
  \quad 
  \sum_{c=1}^{N_c} \left[b_c\,\ov{b}_c - \widetilde{b}_c\, 
  \ov{\widetilde{b}}_c
   \right]=0\;.
}
The first two can be interpreted as the F-term constraints 
$\mbox{Re~}b_c\widetilde{b}_c=0$ and $\mbox{Im~}b_c\widetilde{b}_c=0$, the 
last one as the D-term constraint $\sum_{c=1}^{N_c}|b_c|^2-|\widetilde{b}_c|^2=0$.

As seen in (\ref{zeromm}), there are only $2N_c$ fermionic zero modes. 
However, in \cite{Billo:2002hm} it was shown in the T-dual picture that they have to satisfy the ADHM constraints
\eq{
  \label{adhm_ferm}
  \sum_{c=1}^{N_c} \left[ \widetilde{\beta}_c\: b_c + \beta_c\: 
  \widetilde{b}_c 
  \right]=0 
  \qquad\mbox{and}\qquad
  \sum_{c=1}^{N_c} \left[\widetilde{\beta}_c\: \ov{\widetilde{b}}_c - 
  \beta_c\: 
  \ov{b}_c\right]=0\;.
}
These constraints can be recovered in the present setup by performing the integration over the aforementioned zero modes $\ov\theta_i$ explicitly \cite{Billo:2002hm,Argurio:2007vq}. More concretely, one can absorb the instanton zero modes with terms like $\tilde{\beta}_c\, b_c\, \ov \theta$ and integrate over
$\ov \theta$ to arrive at a $\delta$-function realisation of \eqref{adhm_ferm}.

All these constraints, i.e. equations \eqref{adhm_bos} and \eqref{adhm_ferm}, have to be implemented in the general formula for the superpotential \eqref{general_w}. We will not perform the $\ov \theta$ integration explicitely but realise all the constraints by the means of delta functions. This ensures the gauge invariance of the integration measure.

%%%%%%%%%%%%%%%%%%%%%%%%%%%%%%%%%%%%%%%%%%%%%%%
%%%%%%%%%%%%%%%%%%%%%%%%%%%%%%%%%%%%%%%%%%%%%%%

\paragraph{E2--D6$_f$:} In this sector, the ground state energy in the NS sector is positive, so that there are no bosonic zero modes. One only gets $N_f$ pairs of non-chiral $\lambda_f, \widetilde\lambda_f$ zero modes from eq. (\ref{zeronn}) yielding the measure 
\bea
      \int \prod_{f=1}^{N_f}  d\lambda_f\, d\widetilde{\lambda}_f\; .
\eea

%%%%%%%%%%%%%%%%%%%%%%%%%%%%%%%%%%%%%%%%%%%%%%%
%%%%%%%%%%%%%%%%%%%%%%%%%%%%%%%%%%%%%%%%%%%%%%%

To summarise, the total integration measure for the computation of the E2-instanton generated superpotential \eqref{general_w} reads
\eq{
  \label{our_measure}
  \int d^4x&\, d^2\theta\,\, \prod_{c=1}^{N_c}  
  db_c\,d\ov{b}_c\, d\widetilde b_c\, d\ov{\widetilde b}_c   
  d\beta_c\,d\ov{\beta}_c\,\,  
  \prod_{f=1}^{N_f}  d\lambda_f\, d\widetilde{\lambda}_f\ \,
  \delta^F( \widetilde{\beta}_c\: b_c + \beta_c\: \widetilde{b}_c) \\
  &\delta^F(\widetilde{\beta}_c\: \ov{\widetilde{b}}_c - \beta_c\: 
  \ov{b}_c)\,\, 
  \delta^B( b_c\,\widetilde{b}_c+\ov{b}_c\,\ov{\widetilde{b}}_c )\,\,
  \delta^B( b_c\,\widetilde{b}_c-\ov{b}_c\,\ov{\widetilde{b}}_c)\,\,
  \delta^B( b_c\,\ov{b}_c - \widetilde{b}_c\, \ov{\widetilde{b}}_c )
  \;,
}
where in the two fermionic and three bosonic ADHM constraints summation
over the colour index is understood. It is rather amusing that the
open string zero modes precisely give the flat ADHM measure
with the quadratic constraints arising from the F-term and 
D-term constraints for the effective gauge theory on the
E2-instanton.

%%%%%%%%%%%%%%%%%%%%%%%%%%%%%%%%%%%%%%%%%%%%%%%
%%%%%%%%%%%%%%%%%%%%%%%%%%%%%%%%%%%%%%%%%%%%%%%
%%%%%%%%%%%%%%%%%%%%%%%%%%%%%%%%%%%%%%%%%%%%%%%
%%%%%%%%%%%%%%%%%%%%%%%%%%%%%%%%%%%%%%%%%%%%%%%

\subsection{The field theory limit}
\label{secfield}

Now let us discuss the aforementioned field theory limit in more detail.
We have a stack of colour D6-branes wrapping a three-cycle
of size 
\bea
v_c={V_c\over \ell_s^3}\cong {R_c^3\over (\alpha')^{3\over 2}}=r_c^3
\eea
in string units. The flavour branes wrap a different cycle
with volume $v_f= r_c\, r_T^2$, where $T$ stands for transversal.\footnote{We are thinking here of the situation on the six-torus, where  the flavour- and the colour-cycle share one common direction.}
The volume of the entire CY in string units is $v_{\rm CY}\cong r^3_c\, r^3_T$ and we have two dimensionless scales $r_c, r_T$ and one length scale $\sqrt{\alpha'}$.

The effective four-dimensional couplings in our configuration are
the Yang--Mills coupling of the colour branes
\bea
     {4\pi\over g_{c}^2}= {1\over g_s}{V_c\over \ell_s^3}\cong
     {r_c^3 \over g_s}\;,
\eea
the Yang--Mills coupling of the flavour branes
\bea
     {4\pi \over g_{f}^2}= {1\over g_s}{V_f\over \ell_s^3}\cong
     {r_c\, r_T^2 \over g_s}
\eea
and the Planck mass 
\bea
     M_{\rm Pl}^2\cong{1 \over (2\pi\alpha')\,  g_s^2} r_c^3\, r_T^3 \;.
\eea
Clearly, the gauge coupling of the colour branes 
should be fixed at finite $g_c$. In order
for the Dirac--Born--Infeld theory on these branes
to reduce to the Yang--Mills Lagrangian we have
to suppress all higher $\alpha'$ corrections, which
means we have to take the $\alpha'\to 0$ limit.

On the other hand the gauge theory on the flavour branes
should decouple from the dynamics, which means that
$g_f\to 0$ or, in other words, that the transverse space of volume $v_T=r_T^3$
should decompactify, i.e. $r_T\to \infty$.
In this limit however, also $M_{\rm Pl}\to \infty$ so that
gravity decouples.

In this limit the world-sheet instantons, giving stringy corrections
to the disc correlators, scale like
\bea
  \exp\left( -{J\over \alpha'}\right)\simeq \exp(- r_T^2 ) \to 0 \;,
\eea
so that only trivial instantons of zero size contribute to
the field theory amplitude.
Moreover, in the one-loop Pfaffian all the massive 
string modes decouple in the $\alpha'\to 0$ limit.
The same holds for potential Kaluza-Klein and winding modes
in the various E2--D6 brane sectors, as their masses
scale like
\bea
  M^2_{\rm KK}={1\over R_c^2}\simeq {1\over \alpha' r_c^2} \to \infty
  \qquad\mbox{and}\qquad
  M^2_{\rm wind}={R_f^2\over {\alpha'}^2}\simeq {r^2_T\over \alpha' } 
  \to  \infty \;.
\eea 
The gauge coupling of the effective zero-dimensional gauge theory
on the E2-brane reads
\bea
  {4\pi\over g_E^2}={\ell_s^4\over g_s}{V_3\over \ell_s^3}\simeq
  {{\alpha'}^2\, r_c^3 \over g_s}\to 0 \;,
\eea
i.e. naively, in the field theory limit the effective theory on the
instanton decouples. However, this theory should provide the 
ADHM constraints, so that it should better not decouple completely.
In fact, as shown in \cite{Billo:2002hm}, the operators up to dimension four
in the E2-action  survive precisely due to extra factor
of $\sqrt{g_E}\simeq \sqrt{g_s}$ in the vertex operators for
the E2--D6 modes.

To summarise, in the field theory limit,
\bea
  \label{ft_limit}
  \alpha'\to 0\;, \qquad v_T\to \infty\;, \qquad v_c={\rm finite}\;,
\eea
the complete string theory instanton amplitude simplifies
dramatically and only the leading order world-sheet
disc instantons and massless states in the 1-loop amplitudes
contribute. Moreover, all higher order $\alpha'$ corrections
in the matter fields are suppressed as well and the effective
theory on the ${\rm E2}$-instanton provides the ADHM constraints
for the massless ${\rm E2}$-${\rm D6}_c$ open string modes.

%%%%%%%%%%%%%%%%%%%%%%%%%%%%%%%%%%%%%%%%%%%%%%%
%%%%%%%%%%%%%%%%%%%%%%%%%%%%%%%%%%%%%%%%%%%%%%%
%%%%%%%%%%%%%%%%%%%%%%%%%%%%%%%%%%%%%%%%%%%%%%%
%%%%%%%%%%%%%%%%%%%%%%%%%%%%%%%%%%%%%%%%%%%%%%%
%%%%%%%%%%%%%%%%%%%%%%%%%%%%%%%%%%%%%%%%%%%%%%%
%%%%%%%%%%%%%%%%%%%%%%%%%%%%%%%%%%%%%%%%%%%%%%%
%%%%%%%%%%%%%%%%%%%%%%%%%%%%%%%%%%%%%%%%%%%%%%%
%%%%%%%%%%%%%%%%%%%%%%%%%%%%%%%%%%%%%%%%%%%%%%%

\section{Disc amplitudes}

In order to evaluate the superpotential (\ref{general_w}) for our locally
engineered configuration of intersecting D6-branes, it essentially
remains to compute the appearing disc diagrams with two fermionic
respectively bosonic zero modes attached. Looking at the extended
quiver diagram in figure \ref{figa}, one realises that, due to the
non-chirality of SQCD, numerous insertions 
of chains of matter fields $\Phi$ and $\widetilde\Phi$ are possible.
However, in the field theory limit only the leading order term
with the minimal number of $\Phi$ and $\widetilde\Phi$ insertions survives.

In this section we compute these  relevant  
one- and two-point CFT disc amplitudes
with two fermionic/bosonic zero modes in the background.
To eventually compare our result to field theory, we first have
to ensure that the string amplitudes and the vertex operators 
are normalised correctly. Here we will explicitly take care
of factors of $g_s$, $\alpha'$ and $v_c$. As we will see at the end
of the computation, additional (often convention dependent) 
numerical factors can always be absorbed into
the definition of the dynamically generated scale $\Lambda$ and are
therefore not taken care of explicitly.

%%%%%%%%%%%%%%%%%%%%%%%%%%%%%%%%%%%%%%%%%%%%%%%
%%%%%%%%%%%%%%%%%%%%%%%%%%%%%%%%%%%%%%%%%%%%%%%
%%%%%%%%%%%%%%%%%%%%%%%%%%%%%%%%%%%%%%%%%%%%%%%
%%%%%%%%%%%%%%%%%%%%%%%%%%%%%%%%%%%%%%%%%%%%%%%

\subsection{Normalisation}

In general, conformal field theory amplitudes carry a normalisation factor $C$, i.e.
\bea
  \langle\langle\, \prod_{k=1}^K  \Phi_{k} \,\rangle\rangle
     = C\, \int {dz_1\ldots dz_{K}\over V_{1,2,K}}\, 
              \langle V_{\Phi_1}(z_1)\, V_{\Phi_{2}}(z_2)\,
               \ldots V_{\Phi_K}(z_K)\, \rangle  \;,
\eea    
where $C$ depends on the boundaries of the disc.
If there is only a single boundary brane D$p$, then $C$ is nothing else
than the tension of the D$p$-brane 
\bea
  C={\mu_p\over g_s}={2\pi\over g_s\, \ell_s^{p+1}}\;.
\eea
However, if the disc contains more boundaries the normalisation is given by
the common sub-locus of all boundaries. For the relevant 
branes D$6_c$, D$6_f$ and
E$2$ the resulting normalisation factors are shown in table \ref{tcs}.
\begin{table}[h]
\centering
\begin{tabular}{|c||c|c|c|c|c|c|c|}
\hline
branes & D$6_c$ & D$6_f$ & D$6_c,\mbox{D}6_f$ & E$2$ & E$2,\mbox{D}6_c$ & E$2,\mbox{D}6_f$ & E$2,\mbox{D}6_c,\mbox{D}6_f$ \\
\hline
$C$ & ${2\pi v_c \over g_s \ell_s^4}$ & ${2\pi v_f \over g_s \ell_s^4}$ 
& ${2\pi \over g_s \ell_s^4}$ & ${2\pi v_c \over g_s }$ 
& ${2\pi v_c \over g_s }$ & ${2\pi  \over g_s }$ & ${2\pi  \over g_s }$ \\ 
\hline
\end{tabular}
\caption{Disc normalisation factors $C$ for the various combinations of possible boundaries. \label{tcs}}
\end{table}

Having fixed the normalisation factors of the disc, we have to take care of the normalisation of the vertex operators. These can be fixed by comparison with the standard normalisation in field theory. Let us first consider the vertex operators for the gauge boson in the $(-1)$-ghost picture
\bea
  V^{(-1)}_{A}(z)=A^\mu \ e^{-\varphi (z)}\,\, \psi_\mu (z)\,\, 
  e^{i p\cdot  X  (z)} \;.
\eea
In this normalisation the space-time field $A^\mu$ is dimensionless.
The standard field theory normalisation is then obtained by scaling $A^\mu=(2\pi\alpha')^{1\over 2}\, A_{\rm phys}^\mu$. The Yang--Mills gauge coupling at the string scale is given by 
\bea
   {1\over g^2_{\rm YM}}={1\over 2}\, C_{{\rm D}6_c} (2\pi\alpha')^2.
\eea
In order not to overload the notation with  too many scaling factors,
in the following we 
first normalise the vertex operators such that the space-time
fields do not carry any scaling dimension. Only in the very end
we move to physical fields by introducing appropriate
$(2\pi\alpha')^{\Delta/2}$ factors, where $\Delta$ denotes the
four-dimensional scaling dimension of the field.

Next we have to normalise the matter fields localised on the intersection
of the two D6-branes. We want the kinetic term to be normalised
as in the work by Affleck, Dine and Seiberg 
\bea
  S_{\rm matter}=\int dx^4 {1\over g^2_{\rm YM}}\, \partial_\mu \Phi\, 
  \partial^\mu \ov{\Phi} \;.  
\eea
Since now the disc normalisation does not contain the volume $v_c$, 
we have to put this factor in the normalisation of the vertex operators
leading to   
\bea
  V^{(-1)}_{\Phi}(z)=\sqrt{v_c}\,\, \Phi\ e^{-\varphi (z)}\,\, 
  \Sigma^{cf}_{1/2}(z)\,\, e^{i p\cdot  X  (z)} \;,
\eea
where $\Sigma^{cf}_{1/2}$ is an internal  
twist operator of conformal dimension $h=1/2$.

As explained in section \ref{seca}, 
for the instanton zero modes not to decouple
in the field theory limit, their vertex operators must contain extra
factors of $\sqrt{g_s}$. This leads to the following form of the
vertex operator for the E2--D6$_c$ bosonic zero modes $b_c$
\bea
  V^{(-1)}_{b_c}(z)= \sqrt{g_s\over 2\pi v_c}\, \,b_c\ 
  e^{-{\varphi(z)}}\, \prod_{\mu=0}^3 \sigma^\mu_{1\over 16}(z) 
  \, \prod_{\nu=0}^3 s^\nu_{1\over 16}(z)\;,
\eea
where the $\sigma^\mu_{1\over 16}(z)$ denote the bosonic twist fields of conformal dimension
$h=1/16$ and  $s^\nu_{1\over 16}(z)$ the fermionic ones of the same conformal dimension. 
Similarly, for the fermionic instanton zero modes in this sector we write
\bea
   V^{(-1/2)}_{\beta_c}(z)=\sqrt{g_s\over 2\pi v_c}\,\, \beta_{c}\ 
   e^{-{\varphi(z)\over 2}}  \prod_{\mu=0}^3 \sigma^\mu_{1\over 
   16}(z)\, \, 
   \Sigma^c_{3/8}(z)\;.                   
\eea
Taking into account the different disc normalisations for the 
charged matter fermionic zero modes, the vertex operator is
\bea
   V^{(-1/2)}_{\lambda_f}(z)=\sqrt{g_s\over 2\pi}\, \,\lambda_{f}\ 
   e^{-{\varphi(z)\over 2}}\, \prod_{\mu=0}^3 \sigma^\mu_{1\over 16}(z)
   \, \, \Sigma^f_{3/8}(z)\;.
\eea
In the fermionic vertex operators, $\Sigma_{3/8}$ denote 
appropriate Ramond fields, whose form depends on the concrete
CFT describing of the internal Calabi--Yau manifold.

%%%%%%%%%%%%%%%%%%%%%%%%%%%%%%%%%%%%%%%%%%%%%%%
%%%%%%%%%%%%%%%%%%%%%%%%%%%%%%%%%%%%%%%%%%%%%%%
%%%%%%%%%%%%%%%%%%%%%%%%%%%%%%%%%%%%%%%%%%%%%%%
%%%%%%%%%%%%%%%%%%%%%%%%%%%%%%%%%%%%%%%%%%%%%%%

\subsection{Three-point amplitudes}
\label{sec_three_point}

As we mentioned earlier, in the field theory limit \eqref{ft_limit} the computation of the single 
instanton generated superpotential \eqref{general_w} simplifies considerably. In particular, for our setup shown in figure \ref{figa} we are left with only two disc diagrams with insertion of two fermionic zero modes from the 
set $\beta_c,\widetilde\beta_c,\lambda_f, \widetilde\lambda_f$. 
See for instance \cite{Cremades:2003qj,Cvetic:2003ch,Abel:2003vv,Lust:2004cx,Bertolini:2005qh} for the conformal field
theory computation of disc amplitudes for intersecting D-brane models.  
The appropriate combination of these zero modes can be obtained by looking at the 
extended quiver diagram in figure \ref{figa}. However, because of $U(1)$ charge cancellation on the world-sheet, the fermionic zero modes couple to anti-holomorphic matter fields.
The explicit form of the first disc correlator is
\bea
  \langle \langle\, \ov{\widetilde{\Phi}}_{cf}  \,\rangle\rangle^{\rm
  disc}_{\beta_c\, \widetilde{\lambda}_f}={2\pi\over g_s}\,\int 
  {dz_1\,dz_2\,dz_3 \over V_{123} } \,\,
  \langle V^{(-1/2)} _{\beta_c}(z_1)\, V^{(-1)}_{ 
  \ov{\widetilde{\Phi}}_{cf}}(z_2)\, 
  V^{(-1/2)}_{\widetilde{\lambda}_f}(z_3)\rangle \;.
\eea 
The second disc correlator is $ \langle \langle\, \ov{\Phi}_{fc}  \,\rangle\rangle^{\rm disc}_{\lambda_f\, \widetilde\beta_c}$ and has a form analogous to the first one.

The evaluation of these two disc correlators is easily achieved using the following three-point functions
\eqa{l@{\hspace{5pt}}c@{\hspace{5pt}}l}{
   {\displaystyle\langle \sigma^\mu_{1\over 16}(z_1)\, 
     \sigma^\nu_{1\over 16}(z_3)\rangle} 
   &=& {\displaystyle
     {\delta^{\mu\nu}\over (z_1-z_3)^{1\over 8}}}\;, \\
   {\displaystyle
   \langle e^{-{\varphi(z_1)\over 2}}\, e^{-\varphi (z_2)}\, 
     e^{-{\varphi(z_3)\over 2}}\rangle}
   &=& {\displaystyle{1\over (z_1-z_2)^{1\over 2}\,(z_1-z_3)^
     {1\over 4}\,(z_2-z_3)^{1\over 2} }}\;, \\ 
   {\displaystyle
   \langle \Sigma^c_{3\over 8}(z_1)\, 
     \ov{\widetilde{\Sigma}}^{cf}_{1\over 2}(z_2)\, 
     \Sigma^f_{3\over 8}(z_3) \rangle}
   &=&{\displaystyle
     {1\over (z_1-z_2)^{1\over 2}\,(z_1-z_3)^{1\over 4}\,(z_2-z_3)^
     {1\over 2} }\;, }
} 
and we directly obtain the effective term in the instanton action
\bea
   \label{lag_ferm}
   {\cal L}_{\rm ferm} =  \beta_c\, \ov{\widetilde{\Phi}}_{cf} \, 
   \widetilde{\lambda}_f
   + \lambda_f\, \ov{\Phi}_{fc} \widetilde\beta_c \;. 
\eea
Note that this is the same Lagrangian as appears in the field theory calculation
\cite{Dorey:2002ik}.

%%%%%%%%%%%%%%%%%%%%%%%%%%%%%%%%%%%%%%%%%%%%%%%
%%%%%%%%%%%%%%%%%%%%%%%%%%%%%%%%%%%%%%%%%%%%%%%
%%%%%%%%%%%%%%%%%%%%%%%%%%%%%%%%%%%%%%%%%%%%%%%
%%%%%%%%%%%%%%%%%%%%%%%%%%%%%%%%%%%%%%%%%%%%%%%

\subsection{Four-point amplitude}
\label{sec_four_point}

\begin{figure}[t]
\begin{center}

\setlength{\unitlength}{0.75pt}
\begin{picture}(500,160)(-250,20)

\thicklines\multiput(-158,100)(10,0){10}{\line(1,0){6}}
\thicklines\put(-160,100){\line(-1,1){50}}
\thicklines\put(-160,100){\line(-1,-1){50}}
\thicklines\put(-60,100){\line(1,1){50}}
\thicklines\put(-60,100){\line(1,-1){50}}

\put(-235,150){$\Phi$}
\put(-235,40){$\ov{\Phi}$}
\put(15,150){$b$}
\put(15,40){$\ov{b}$}
\put(-115,65){${\displaystyle \frac{1}{p^2}}$}

\thicklines\put(110,150){\line(1,-1){100}}
\thicklines\put(110,50){\line(1,1){100}}
\thicklines\put(160,100){\circle*{8}}

\put(85,150){$\Phi$}
\put(85,40){$\ov{\Phi}$}
\put(225,150){$b$}
\put(225,40){$\ov{b}$}

\end{picture}
\caption{Four-point massless string exchange and four-point vertex interaction. \label{figfour}}
\end{center}
\end{figure}
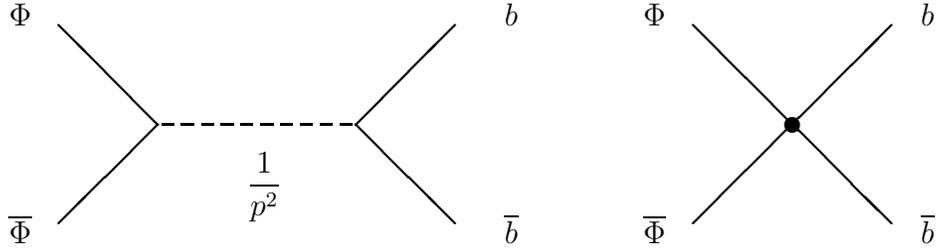

In the field theory limit we essentially have to compute one disc amplitude with insertion of two bosonic zero modes. Looking again at the extended quiver in figure \ref{figa}, we are led to insert the bosonic zero modes $\{b,\ov{b}\}$ and $\{\ov{\widetilde{b}},\widetilde{b}\}$.%%
%%%%%%%%%%%%%%%%%%%%
%%%%%%%%%%%%%%%%%%%%
\footnote{Naively, it is also possible to pair the zero modes as $\{b,\widetilde{b}\}$ and $\{\ov{\widetilde{b}},\ov{b}\}$. However, let us for the moment go back to the notation $b_{\alpha,c}$ from the beginning and attach labels $\alpha$ also to the spin fields $s_{\frac{1}{16}}^{\alpha}$ of equation \eqref{bos_two_point}. Equation (A.19) in \cite{Billo:2002hm} tells us that $\langle s^{\alpha}_{\frac{1}{16}}(z) s^{\beta}_{\frac{1}{16}}(w)\rangle\sim -\frac{\epsilon^{\alpha\beta}}{(z-w)^{1/2}}$ and therefore only the insertion of $\{b_{1,c},\ov{b}_{2,c}\}$ or $\{b_{2,c},\ov{b}_{1,c}\}$ into the disc amplitude gives a non-vanishing result. Converting this back into our notation, this is precisely the choice we mentioned.}
%%%%%%%%%%%%%%%%%%%%
%%%%%%%%%%%%%%%%%%%%
Because of $U(1)$ charge cancellation on the world-sheet, the matter fields can only be paired as $\Phi\,\ov{\Phi}$ and as $\widetilde{\ov{\Phi}}\,\widetilde{\Phi}$. The explicit form of the CFT four-point function is the following
\eq{
  \label{fourpoint_amp}
  \langle\langle\, \Phi_{cf} \ov{\Phi}_{fc'}\,\rangle 
  \rangle^{\rm disc}_{b_c\, \ov b_{c'}}
  ={2\pi\over g_s}\, \int {dz_1\,dz_2\,dz_3\,dz_4 \over V_{124} } 
  \,\,\langle V^{(-1)} _{b_c}(z_1)\, V^{(0)}_{\Phi_{cf}}(z_2)\, 
  V^{(0)}_{\ov{\Phi}_{fc'}}(z_3)\, 
  V^{(-1)}_{\ov b_{c'}}(z_4)\rangle \;,
} 
and similarly for $\{\ov{\widetilde{b}},\widetilde{b}\}$ and $\widetilde{\ov{\Phi}}\,\widetilde{\Phi}$. The vertex operators for the bosons of the matter superfields in the $(0)$-ghost picture are given by
\bea
  V^{(0)}_{\Phi}(z)= \sqrt{v_c}\,\Phi\ \left( \Sigma_{1}(z) + i 
  (p \cdot  \psi)\,\Sigma_{1/2}(z)\right) \, e^{i p \cdot X (z)}\;,
\eea
where $\Sigma_{1}(z)=\{G_{-1/2}, \Sigma_{1/2}(z)\}$ is an internal  operator of conformal dimension $h=1$. The four-point function in the ghost sector and the internal sector simply
reduce to two-point functions
\eqa{@{}l@{\hspace{5pt}}c@{\hspace{5pt}}l@{\hspace{20pt}}l@{\hspace{5pt}}c@{\hspace{5pt}}l@{}}{
  \label{bos_two_point}
  {\displaystyle\langle e^{-{\varphi(z_1)}}\, e^{-{\varphi(z_4)}} 
  \rangle}
  &=& {\displaystyle{1\over (z_1-z_4) }\;,}
  &{\displaystyle
  \langle s^\mu_{1\over 16}(z_1)\, s^\nu_{1\over 16}(z_4) \rangle}
  &=& {\displaystyle{\delta^{\mu\nu}\over (z_1-z_4)^{1\over 8} }\;,} \\
  {\displaystyle\langle \Sigma_{1/2}(z_2) \, 
  \ov{\Sigma}_{1/2}(z_3)  \rangle}
  &=& {\displaystyle{1\over (z_2-z_3) }\;,}
  &{\displaystyle\langle \Sigma_{1}(z_2) \,\ov{\Sigma}_{1}(z_3)\rangle}
  &=&{\displaystyle {1\over (z_2-z_3)^2 }\;.}
}
What remains to be computed are the four-point functions
\eq{
  \langle \sigma^\mu_{1\over 16}(\infty) \, e^{i p_\mu X^\mu (z_2)}\, 
  e^{-i p_\mu X^\mu (z_3)} \, \sigma^\mu_{1\over 16}(0) \rangle
}
and
\eq{  
  \langle s^\mu_{1\over 16}(\infty) \, \psi_\mu(z_2)\, \psi_\mu(z_3) \, 
  s^\mu_{1\over 16}(0) \rangle
}
(no sum over $\mu$). However, these functions are nothing else than the respective two-point functions in the $\mbb Z_2$ twisted sector, which can be obtained from 
\bea
  \langle  X^\mu (z) \, X^\nu (w)\rangle_T=-\eta^{\mu\nu}\, \ln \left[ 
  {z-w\over (\sqrt{z}+\sqrt{w})^2} \right]
\eea
and
\bea
  \langle  \psi^\mu (z) \, \psi^\nu (w)\rangle_T= 
  {\eta^{\mu\nu}\over 2}\, 
  {\sqrt{z\over w}+\sqrt{w\over z}\over z-w}\;.
\eea
Combining all these terms and substituting $x^2=z_3$, the four-point function \eqref{fourpoint_amp} can be expressed as
\bea
   \langle\langle\, \Phi_{cf} \ov{\Phi}_{fc'} \,\rangle
   \rangle^{\rm disc}_{b_c\, \ov b_{c'}}
   =b_c\, \Phi_{cf}\, \ov{\Phi}_{fc'}\, \ov b_{c'} \,\,
   \int_{0}^1 dx\,\, {(1+x)^{p^2-2} \over (1-x)^{p^2+2} }\, 
   \left( 2x+p^2\, (x^2+1) \right)\;.
\eea
The integral is elementary and one finally obtains
\bea
   \langle\langle\,  \Phi \ov{\Phi} \,\rangle\rangle^{\rm
   disc}_{b_c\, \ov b_c}=b_c\, \Phi\, \ov{\Phi}\, \ov b_c \left[
    2^{p^2-1}\, \lim_{x\to 1}\, (1-x)^{-p^2-1}
    -{1\over 2} \right] \;.
\eea
This four-point string amplitude corresponds to the two field theory diagrams shown in figure~\ref{figfour}. The first term, corresponding to the first diagram, is divergent for $p^2\to0$, i.e. on-shell, and reflects the singularity due to the exchange of massless states in the four-point amplitude. The second term corresponds to the second diagram and is momentum-independent. Therefore, it should be interpreted as the local four-point interaction.

In summary, taking also into account the insertion of the bosonic zero modes $\{\ov{\widetilde{b}},\widetilde{b}\}$ and the matter field pairings $\ov{\widetilde{\Phi}}\,\widetilde{\Phi}$, we obtain the effective term in the instanton action originating from bosonic zero modes as
\bea
  {\cal L}_{\rm bos} =  \textstyle{\frac{1}{2}}\;
  b_c\left( \Phi_{cf}\, 
  \ov{\Phi}_{fc'} + \widetilde{\ov{\Phi}}_{cf}\, 
  \widetilde{\Phi}_{fc'} \right) \ov b_{c'} 
  + \textstyle{\frac{1}{2}}\;\ov{\widetilde b}_c \left( \Phi_{cf}\, 
  \ov{\Phi}_{fc'} 
  + \widetilde{\ov{\Phi}}_{cf}\, \widetilde{\Phi}_{fc'} \right) 
  \widetilde b_{c'} \;.
\eea
This again agrees with the field theory Lagrangian \cite{Dorey:2002ik}.

%%%%%%%%%%%%%%%%%%%%%%%%%%%%%%%%%%%%%%%%%%%%%%%
%%%%%%%%%%%%%%%%%%%%%%%%%%%%%%%%%%%%%%%%%%%%%%%
%%%%%%%%%%%%%%%%%%%%%%%%%%%%%%%%%%%%%%%%%%%%%%%
%%%%%%%%%%%%%%%%%%%%%%%%%%%%%%%%%%%%%%%%%%%%%%%
%%%%%%%%%%%%%%%%%%%%%%%%%%%%%%%%%%%%%%%%%%%%%%%
%%%%%%%%%%%%%%%%%%%%%%%%%%%%%%%%%%%%%%%%%%%%%%%
%%%%%%%%%%%%%%%%%%%%%%%%%%%%%%%%%%%%%%%%%%%%%%%
%%%%%%%%%%%%%%%%%%%%%%%%%%%%%%%%%%%%%%%%%%%%%%%

\section{The Affleck--Dine--Seiberg Superpotential}

Having engineered a local intersecting
D6-brane model giving rise to $SU(N_c)$ super Yang--Mills
theory with $N_f$ non-chiral flavours and having determined
the leading order disc diagrams, we now come to the evaluation
of the zero mode integrals in (\ref{general_w}) to see whether one can really
recover the Affleck--Dine--Seiberg (ADS)  superpotential
\bea
  W=   { \Lambda^{3N_c-N_f}\over {\rm det}(M_{ff'}) } \;.
\eea 
Note that here the matter fields carry their canonical scaling dimension
$\Delta=1$. The dynamically generated scale $\Lambda$ is defined as
\bea
  \label{dynlam}
  \left( {\Lambda\over \mu}\right)^{3N_c-N_f} = \exp \left( 
  {-{8\pi^2\over g^2_{c}(\mu)}}\right) \;.
\eea 
Performing the corresponding single E2-instanton computation
in the local D-brane model, we are aiming to explicitly derive the
ADS superpotential in the field theory limit.
Since we think that our methods are applicable to more
general instanton computations, we provide the
actual evaluation of the zero mode integrals in quite some
detail.

%%%%%%%%%%%%%%%%%%%%%%%%%%%%%%%%%%%%%%%%%%%%%%%
%%%%%%%%%%%%%%%%%%%%%%%%%%%%%%%%%%%%%%%%%%%%%%%
%%%%%%%%%%%%%%%%%%%%%%%%%%%%%%%%%%%%%%%%%%%%%%%
%%%%%%%%%%%%%%%%%%%%%%%%%%%%%%%%%%%%%%%%%%%%%%%
%%%%%%%%%%%%%%%%%%%%%%%%%%%%%%%%%%%%%%%%%%%%%%%
%%%%%%%%%%%%%%%%%%%%%%%%%%%%%%%%%%%%%%%%%%%%%%%
%%%%%%%%%%%%%%%%%%%%%%%%%%%%%%%%%%%%%%%%%%%%%%%
%%%%%%%%%%%%%%%%%%%%%%%%%%%%%%%%%%%%%%%%%%%%%%%

\subsection{Fermionic zero mode integration}

In this subsection we compute to lowest order in $\alpha'$ the contribution of the fermionic zero modes to the E$2$-instanton generated superpotential \eqref{general_w}. The relevant disc diagrams have been computed in section~\ref{sec_three_point}. However, since we will perform an integration over instanton zero modes, we should replace the matter fields $\Phi_{cf}$ by their VEVs $\langle\Phi_{cf}\rangle$ and similarly for $\widetilde{\Phi}_{fc}$ and their conjugates. Furthermore, we have to satisfy the D-term constraint for the matter fields to ensure the supersymmetry of our setup so we can apply formula \eqref{general_w}. Specifically, the D-term constraint implies
\eq{
  \label{d_term}
  \langle\Phi_{cf}\rangle = \langle\ov{\widetilde{\Phi}}_{cf}\rangle
  \qquad\mbox{and}\qquad
  \langle\widetilde{\Phi}_{cf}\rangle = \langle\ov{\Phi}_{cf}\rangle \;.
}
Using these relations, one can rewrite \eqref{lag_ferm} to obtain the relevant three-point couplings  
\bea
   \beta_c\, \langle \Phi_{cf}\rangle \, 
   \widetilde{\lambda}_f
   + \lambda_f\, \langle \widetilde{\Phi}_{fc}\rangle
   \widetilde\beta_c 
\eea
where $c=1,\ldots, N_c$ and $f=1,\ldots, N_f$. The sum over repeated indices is understood and in the following we will omit the VEV brackets for ease of notation.

The next step is to implement the fermionic ADHM constraints \eqref{adhm_ferm} in terms of delta-functions. For Grassmann-variables they read 
\begin{equation}
        \delta\left(\widetilde{\beta}_{c_1}\:b_{c_1}+
                \beta_{c_2}\:\widetilde{b}_{c_2}\right)\:
        \delta\left(\widetilde{\beta}_{c_3}\:\bar{\widetilde{b}}_{c_3}-
                \beta_{c_4}\:\bar{b}_{c_4}\right)=
        \left(\widetilde{\beta}_{c_1}\:b_{c_1}+\beta_{c_2}\:\widetilde{b
        }_{c_2}
        \right)
        \left(\widetilde{\beta}_{c_3}\:\bar{\widetilde{b}}_{c_3}-
                \beta_{c_4}\:\bar{b}_{c_4}\right).
\end{equation}
Taking also into account the fermionic part of the integration measure \eqref{our_measure}, we arrive at the following expression for the fermionic contribution to the superpotential 
\begin{multline}
  \label{ferm_int}
  I_{\rm ferm}(\Phi, \widetilde\Phi,b,\widetilde b)=
  \int \prod_{c=1}^{N_c} d\beta_c d\widetilde{\beta}_c
  \prod_{f=1}^{N_f} d\lambda_f d\widetilde{\lambda}_f\;
  \left(
  \widetilde{\beta}_{c_1}b_{c_1} \ov{\widetilde{b}}_{c_2} 
  \widetilde{\beta}_{c_2}
   -\widetilde{\beta}_{c_1}b_{c_1} \ov{b}_{c_2} \beta_{c_2}
  \right. \\ \left.
   +\beta_{c_1}\widetilde{b}_{c_1} \ov{\widetilde{b}}_{c_2} 
   \widetilde{\beta}_{c_2}
   -\beta_{c_1}\widetilde{b}_{c_1} \ov{b}_{c_2} \beta_{c_2}
  \right)\,\,
  \exp\left(\beta_c \Phi_{cf} \widetilde{\lambda}_f +
            \lambda_f \widetilde{\Phi}_{fc} \widetilde{\beta}_c \right) \;.
\end{multline}
Note the dependence not only on the matter fields but also on 
the bosonic zero modes $b,\widetilde b$. The fermionic modes $\beta$ and $\widetilde{\beta}$ in front of the exponent 
determine which part in the series expansion is picked out by the integral. 
For instance, in the case of a prefactor $\widetilde{\beta}\beta$ only 
$(\beta_c \Phi_{cf} \widetilde{\lambda}_f)^{N_c-1}$ and 
$( \lambda_f \widetilde{\Phi}_{fc} \widetilde{\beta}_c )^{N_c-1}$ survive the $d\beta$ integration. This implies, however, that the integral \eqref{ferm_int} can only be non-zero if and only if
\eq{
  N_f=N_c-1 \;.
}
One can check that for prefactors $\beta\beta$ and $\widetilde{\beta}\,\widetilde{\beta}$ the complete fermionic integral vanishes, since the number of modes $\lambda$ is equal the number of modes $\widetilde{\lambda}$. Therefore, we recovered nicely the constraint of  Affleck, Dine and Seiberg for the instanton generated
ADS superpotential.

By differentiating under the integral sign, and using some of the formulae of appendix \ref{app_gen_form}, one finds that the final result for the fermionic zero mode integration is
\eq{
  \label{int_ferm_01}
   I_{\rm ferm}(\Phi, \widetilde\Phi,b,\widetilde b)=
   \sum_{p,q=1}^{N_c} 
   \left(-1\right)^{p+q}
   \left( b_p\ov{b}_q + \ov{\widetilde{b}}_p \widetilde{b}_q \right)
   \det\bigl[ \Phi\widetilde{\Phi}\bigr\rvert_{q,p}\bigr] \;.
}   
Here, and in the following, $\Phi$ is the $N_c\times (N_c-1)$ matrix with elements $\Phi_{c,f}$ and similarly $\widetilde{\Phi}$ is an $(N_c-1)\times N_c$ matrix. The symbol $A\rvert_{q,p}$ denotes the matrix obtained from $A$ by deleting the $q$'th row and $p$'th column. The $(q,p)$'th element of a matrix
$A$ will be denoted by $A_{q,p}$.

%%%%%%%%%%%%%%%%%%%%%%%%%%%%%%%%%%%%%%%%%%%%%%%
%%%%%%%%%%%%%%%%%%%%%%%%%%%%%%%%%%%%%%%%%%%%%%%
%%%%%%%%%%%%%%%%%%%%%%%%%%%%%%%%%%%%%%%%%%%%%%%
%%%%%%%%%%%%%%%%%%%%%%%%%%%%%%%%%%%%%%%%%%%%%%%

\subsection{Bosonic zero mode integration}

Let us now turn to the evaluation of the bosonic zero mode integrals in \eqref{our_measure}. The relevant disc diagrams have been computed in section~\ref{sec_four_point}. Following the same reasoning as for the three-point amplitudes, we replace the matter fields by their VEVs and use the D-term constraint \eqref{d_term} to arrive at the four-point couplings
\bea
  \label{loop_bos}
  b_c\, \langle\Phi_{cf}\rangle\,\langle\widetilde{\Phi}_{fc'}
  \rangle \,\ov b_{c'} 
  + \ov{\widetilde b}_c \,\langle\Phi_{cf}\rangle\,
  \langle\widetilde{\Phi}_{fc'} \rangle\,
  \widetilde b_{c'} \;.
\eea 
Note that in the following we will again omit the VEV brackets for ease of notation. The ADHM constraints for the bosonic moduli space, or equivalently the D- and F-term constraints for the E$2$, are given in equation \eqref{adhm_bos}. Again, these have to be implemented as delta-functions in the integration measure \eqref{our_measure}. The explicit form of the bosonic part reads 
\eq{ 
  \label{bos_measure}
  &\int \prod_{c=1}^{N_c} db_c\: d\ov{b}_c\: d\widetilde{b}_c\: 
  d\ov{\widetilde{b}}_c\;
  \delta\left( \ov{b}_c\ov{\widetilde{b}}_c + \widetilde{b}_cb_c \right)
  \delta\left( i\ov{b}_c\ov{\widetilde{b}}_c - i\widetilde{b}_cb_c 
  \right)
  \delta\left( \ov{b}_cb_c - \widetilde{b}_c\ov{\widetilde{b}}_c 
  \right) \\
  =&\int \prod_{c=1}^{N_c} db_c\: d\ov{b}_c\: d\widetilde{b}_c\: 
  d\ov{\widetilde{b}}_c\; \int dk_1
  \exp\left( ik_1 (\ov{b}_c\ov{\widetilde{b}}_c + \widetilde{b}_c b_c) 
  \right) \\
  & \hspace{50pt} \times \int dk_2 \exp\left( 
             -k_2(\ov{b}_c\ov{\widetilde{b}}_c -  \widetilde{b}_c b_c) 
             \right)
        \int dk_3 \exp\left( ik_3(\ov{b}_cb_c - 
        \widetilde{b}_c\ov{\widetilde{b}}_c)
  \right)\, .  
}
Finally, we combine \eqref{loop_bos}, \eqref{bos_measure} and the bosonic fields from the fermionic integration \eqref{int_ferm_01} into the bosonic integral
\eq{
  &\int d^3k \int \prod_{c=1}^{N_c} db_c\: d\ov{\widetilde{b}}_c\: 
  d\ov{b}_c\:d\widetilde{b}_c
  \left( b_p\ov{b}_q + \ov{\widetilde{b}}_p \widetilde{b}_q \right)
  \exp\left( -
  \biggl[\begin{array}{c} b \\ \ov{\widetilde{b}} \end{array}\biggl]^T
  \biggl[\begin{array}{cc} M_{(1)} & M_{(2)} \\ M_{(3)} & M_{(4)} 
         \end{array}\biggr]
  \biggl[\begin{array}{c} \ov{b} \\ \widetilde{b} \end{array}\biggl]
  \right), 
  \label{int_bos_01}
}
where the $b$'s are vectors with the $N_c$ entries $b_c$ and the $M$'s are 
$N_c\times N_c$ matrices of the following form
\eqa{l@{\hspace{40pt}}l}{
  M_{(1)} = -\Phi\widetilde{\Phi} + \epsilon\, \eins  -ik_3\,\eins\;, &
  M_{(2)} = -ik_1\,\eins-k_2\,\eins\;, \\
  M_{(3)} = -ik_1\,\eins+k_2\,\eins\;, &
  M_{(4)} = -\Phi\widetilde{\Phi} + \epsilon\,\eins +ik_3\,\eins\;. 
}   
Here were added an infinitesimal parameter $\epsilon>0$ in order to regularise the complex Gaussian integrals \eqref{int_bos_01}. At the end of the computation we will take the $\epsilon\to 0$ limit. Note that bosonic fields in front of the exponential can be written as
\eq{
   b_p\ov{b}_q + \ov{\widetilde{b}}_p\widetilde{b}_q 
   = -\frac{\partial}{\partial M_{(1)p,q}} 
     -\frac{\partial}{\partial M_{(4)p,q}} \;,
}
where for instance the first derivative is with respect to the $(p,q)$'th 
element of the matrix $M_{(1)}$. Then one can perform the Gaussian integrals in \eqref{int_bos_01} to obtain
\begin{align}
  \label{int_bos_02}
  &\int d^3k \left( -\frac{\partial}{\partial M_{(1)p,q}} 
  -\frac{\partial}{\partial M_{(4)p,q}} \right)\,\, 
  \det \left( \biggl[\begin{array}{cc} M_{(1)} & M_{(2)} \\ M_{(3)} & 
              M_{(4)} \end{array}\biggr] \right)^{-1}
  \\
  \label{int_bos_03}
  =&\int d^3k \left( 
  \left(M_{(1)}^{-1}\right)_{q,p} + \left(M_{(4)}^{-1}\right)_{q,p}
  \right)    \,\,
   \det \left( \biggl[\begin{array}{cc} M_{(1)} & M_{(2)} \\ M_{(3)} & 
              M_{(4)} \end{array}\biggr] \right)^{-1}   \\
  \label{int_bos_04}
  =&-2\int d^3k \sum_{r=1}^{N_c}\frac{ 
  \Bigl( \Phi\widetilde{\Phi} - \epsilon \eins \Bigr)_{q,r}  
  \Bigl( \bigl( \Phi\widetilde{\Phi} -\epsilon\eins \bigr)^2 + k^2\eins 
  \Bigr)^{-1}_{r,p}
  }{
  \det \Bigl[ \bigl( \Phi\widetilde{\Phi} -\epsilon\,\eins \bigr)^2 + 
  k^2\,\eins \Bigr]
  } \;.
\end{align}  
From line \eqref{int_bos_02} to \eqref{int_bos_03} we used formula 
\eqref{app_det_deriv} from the appendix. From \eqref{int_bos_03} to line 
\eqref{int_bos_04} we used \eqref{app_det_inv} and \eqref{app_det_block}. Note that $M_{(1)}^{-1}$ stands for the upper-left block of the block-matrix $M^{-1}$, i.e. it is not the inverse of $M_{(1)}$.

\bigskip
At this point let us summarise the results of the fermionic and bosonic zero mode integration so far. Combining equations \eqref{int_ferm_01} and 
\eqref{int_bos_04} and going to spherical coordinates we find the expression 
\begin{multline}
  \label{int_bos_05}
  I_{\rm bos}(\Phi, \widetilde\Phi)=
  \int d^4x\, d^2\theta \;
  \int_0^{\infty} dk\;k^2 \\  
  \times\sum_{p,q,r=1}^{N_c} \left(-1\right)^{p+q} 
  \frac{
  \det\bigl[ \Phi\widetilde{\Phi}\bigr\rvert_{q,p} \bigr]\,
  \Bigl( \Phi\widetilde{\Phi} - \epsilon\,\eins\Bigr)_{q,r} 
  \Bigl( \bigl( \Phi\widetilde{\Phi} -\epsilon\,\eins \bigr)^2 
  + k^2\,\eins \Bigr)^{-1}_{r,p}
  }
  {\det \Bigl[ \bigl( \Phi\widetilde{\Phi} -\epsilon\,\eins \bigr)^2 
  + k^2\,\eins \Bigr]}
  \;.
\end{multline}
Note that so far still the combination $\Phi\widetilde{\Phi}$ occurs, whereas
eventually we have to get the flavour matrix $\widetilde\Phi {\Phi}$.
In (\ref{int_bos_05}) there is a part which can be simplified using equations \eqref{app_det_sum} and \eqref{app_det_rank}
\eq{
  \sum_{q=1}^{N_c} \left(-1\right)^{p+q} 
 \, \det \bigl[ \Phi\widetilde{\Phi}\bigr\rvert_{q,p} \bigr]\; 
   \Phi\widetilde{\Phi}_{q,r} 
  =\det \bigl[ \Phi\widetilde{\Phi}\bigr] \: \delta_{p,r}
  =0 \;.
}
The remaining part can be rewritten using equations \eqref{app_det_inv02} and \eqref{app_det_sum02} from the appendix  
\eq{
  \label{int_bos_06}
  &\epsilon\int_0^{\infty} dk\;k^2 
  \sum_{p,q=1}^{N_c} \left(-1\right)^{2p+2q}
  \frac{
  \det\bigl[ \Phi\widetilde{\Phi}\bigr\rvert_{q,p}\bigr]\, 
  \det\Bigl[ \Bigl( \bigl( \Phi\widetilde{\Phi} -\epsilon\,\eins 
  \bigr)^2 
  + k^2\,\eins 
  \Bigr)\Bigr\rvert_{p,q}\Bigr]
  }
  {\det \Bigl[ \bigl( \Phi\widetilde{\Phi} -\epsilon\,\eins \bigr)^2 
  + k^2\,\eins \Bigr]^2}
  \\
  =&\epsilon\int_0^{\infty} dk\;k^2 
  \frac{
  \sum_{p=1}^{N_c} \;
  \det\Bigl[  \Bigl( \Phi\widetilde{\Phi} 
       \bigl( \bigl( \Phi\widetilde{\Phi} -\epsilon\,\eins \bigr)^2 
       + k^2\, \eins
       \bigr)\Bigr)\Bigr\rvert_{p,p}\Bigr]
  }
  {\det \Bigl[ \bigl( \Phi\widetilde{\Phi} -\epsilon\,\eins \bigr)^2 
  + k^2\,\eins \Bigr]^2} \;.
}
This expression can be simplified using the results of 
appendix \ref{app_detform} for the numerator and of appendix 
\ref{app_charpol} for the denominator
\eq{
  =&\epsilon\int_0^{\infty} dk\;k^2 
  \frac{\det \Bigl[ \widetilde{\Phi}\Phi 
       \bigl( \bigl( \widetilde{\Phi}\Phi -\epsilon\,\eins \bigr)^2 
       + k^2\,\eins  \bigr)\Bigr]
  }
  {\bigl(ik-\epsilon\bigr)^2\bigl(ik+\epsilon\bigr)^2
  \det \bigl[ \widetilde{\Phi}\Phi -\epsilon\,\eins +ik\,\eins \bigr]^2
  \det \bigl[ \widetilde{\Phi}\Phi -\epsilon\,\eins -ik\,\eins 
  \bigr]^2} \\
  =&\epsilon\int_0^{\infty} dk\;k^2 
  \frac{\det\bigl[ \widetilde{\Phi}\Phi \bigr] }
  {\bigl(ik-\epsilon\bigr)^2\bigl(ik+\epsilon\bigr)^2
  \det \bigl[ \widetilde{\Phi}\Phi -\epsilon\,\eins +ik\,\eins \bigr]
  \det \bigl[ \widetilde{\Phi}\Phi -\epsilon\,\eins -ik\,\eins \bigr]} 
  \;.
}
Finally, let us denote the eigenvalues of the matrix $\widetilde{\Phi}\Phi$ as $\sigma_j$ with $j=1,\ldots, N_c-1$. Then one obtains for equation \eqref{int_bos_05} the following expression
\begin{multline}
  \label{int_bos_07}
  I_{\rm bos}(\Phi, \widetilde\Phi)=
  \int d^4x\, d^2\theta \,\, 
    \det\bigl[ \widetilde{\Phi}\Phi\bigr] \\
  \times\epsilon
  \int_0^{\infty} dk\; 
  \frac{k^2}
  {\bigl(ik-\epsilon\bigr)^2\bigl(ik+\epsilon\bigr)^2
  \prod_{j=1}^{N_c-1}
  \bigl( \sigma_j -\epsilon +ik \bigr)
  \bigl( \sigma_j -\epsilon -ik \bigr)} \;. 
\end{multline}
Let us denote the integrand in the last line as $f(k)$.
Due to the symmetry $k\to -k$ of the integral and the vanishing of the
integrand at infinity, one can rewrite \eqref{int_bos_07} as a contour
integral 
\eq{
 \label{countour}
 I_{\rm bos}(\Phi, \widetilde\Phi)=
 \int d^4x\, d^2\theta \,\, 
 \det\bigl[ \widetilde{\Phi}\Phi\bigr]\, {\epsilon\over 2\pi i}\,\oint 
 f(k)  
}
over the upper half-plane, for instance. The contour integral can be evaluated by virtue of the residue theorem
\bea
  \label{residue_1}
  {\epsilon\over 2\pi i}\,\oint f(k)  =
  \epsilon\:\mbox{Res}\left[ f(k),k=i\epsilon \right] + 
  \epsilon\sum_{j=1}^{N_c-1} \mbox{Res}\left[ f(k),k=i(\sigma_j
  -\epsilon) \right] \;. 
\eea
Generically, all eigenvalues of the matrix $\widetilde{\Phi}\Phi$ are nonzero and one finds for the first term 
\eq{
  \label{residue_2}
  \epsilon\,\mbox{Res}\left[ f(k),k=i\epsilon \right]
  &= -\frac{i}{4}\;\frac{1}{\prod_{j=1}^{N_c-1} \bigl( 
  \sigma_j-2\epsilon\bigr)\sigma_j}
  \left( 1 + \sum_{j=1}^{N_c-1} \frac{2\epsilon^2}{\bigl( 
  \sigma_j-2\epsilon\bigr)\sigma_j}
  \right) \;. 
}
For the calculation of all the other residues in \eqref{residue_1} one can assume 
that $|\sigma_j|>\epsilon$. Therefore, the limit $\epsilon\to0$ can be 
performed before evaluating the residue and so all these terms vanish.
Taking the limit $\epsilon\to 0$ in expression \eqref{residue_2}, the result simplifies drastically and one obtains
\bea
  \label{int_bos_09}
  \lim_{\epsilon\to 0}\,\,  \mbox{Res}\left[ f(k),k=i\epsilon \right] =
   -\frac{i}{4}\;\frac{1}{\prod_{j=1}^{N_c-1} \sigma_j^2 }
   = -\frac{i}{4}\;\frac{1}{\det\bigl[   \widetilde{\Phi}\Phi  \bigr]^2}
   \; .
\eea
Using this result for the contour integration, we arrive at our final result for the bosonic zero mode integration 
\bea 
   I_{\rm bos}(\Phi, \widetilde\Phi)=
   \int d^4 x\, d^2\theta\,\,    
   {1\over  \det\bigl[ \widetilde{\Phi}\Phi \bigr] }  \;.
\eea

\bigskip
The last step to obtain the E2-instanton generated superpotential \eqref{general_w} is to include the contribution of $\exp ( \langle 1 \rangle^{\rm disc} )$
\bea 
   S_W={2\pi^2\over \ell_s^3}\, \int d^4 x\, d^2\theta\,\,    
   {1\over  \det\bigl[ \widetilde{\Phi}\Phi \bigr] } \,   \exp\left(
   -{8\pi^2\over g_c^2(M_s)}\right) 
\eea
with $M_s=(\alpha')^{-1/2}$. 
Transforming to canonically normalised matter fields,
$\Phi=(2\pi\alpha')^{1/2}\, \Phi_{\rm phys}$, we get
\bea 
   S_W= N\, \int d^4 x\, d^2\theta\,\,    
   {M_s^{2N_f+3}\over  \det\bigl[ \widetilde{\Phi}_{phys}\Phi_{phys} 
   \bigr] } \,   \exp\left(
   -{8\pi^2\over g_c^2(M_s)}\right) \;,
\eea
where we have collected all numerical factors appearing during the 
computation into the constant $N$.

Using (\ref{dynlam}) for the dynamically generated scale $\Lambda$,
while absorbing the  numerical factor into $\Lambda$,  
gives precisely the ADS superpotential
\bea
 S_W=\int d^4 x\, d^2\theta \,\,   
        {\Lambda^{3N_c-N_f} \over  {\det}[M_{ff'}] } \;. 
\eea
Therefore, we have derived the ADS superpotential in the field theory limit
from a D-brane instanton. As mentioned, in string theory there will be numerous
corrections to this simple expression. All the massive string states
will appear in the one-loop determinant and there will be an infinite
series of world-sheet instanton corrections to the disc amplitudes.
Moreover, multiple matter field insertions along the boundary of the
disc are possible and give higher $\alpha'$ corrections to the
ADS superpotential. 
Note that these higher order terms will break
some of the global symmetries, like R-symmetry, present
in the field theory limit. 
Therefore, one warning is in order at this point: 
\begin{quote}
Whenever
one simply extends, for instance, string flux superpotentials by
non-perturbative superpotentials from pure field theory, like
the ADS superpotential, one has to ensure that one always remains
in the regime of validity of the field theory limit
of string theory, i.e. at large transversal radii and small VEVs for the 
matter fields, i.e. $\langle \Phi\rangle \ll M_s$. 
\end{quote}
We would also like to add a few methodological remarks concerning our derivation
of the ADS superpotential. Recall that Affleck, Dine and Seiberg inferred the form of their
superpotential solely from considerations of symmetry and then calculated that it is in fact
generated by instantons. Therefore, whenever one has constructed an SQCD theory with $N_f = N_c-1$,
regardless of its origin, it is a priori clear that it should feature an ADS superpotential. In this
regard, our stringy calculation of the ADS superpotential can also be considered a benchmark for
the instanton calculus of \cite{Blumenhagen:2006xt}, showing in a non-trivial way that the prescription
given there leads to the correct, known result in our setup.

%%%%%%%%%%%%%%%%%%%%%%%%%%%%%%%%%%%%%%%%%%%%%%%
%%%%%%%%%%%%%%%%%%%%%%%%%%%%%%%%%%%%%%%%%%%%%%%
%%%%%%%%%%%%%%%%%%%%%%%%%%%%%%%%%%%%%%%%%%%%%%%
%%%%%%%%%%%%%%%%%%%%%%%%%%%%%%%%%%%%%%%%%%%%%%%
%%%%%%%%%%%%%%%%%%%%%%%%%%%%%%%%%%%%%%%%%%%%%%%
%%%%%%%%%%%%%%%%%%%%%%%%%%%%%%%%%%%%%%%%%%%%%%%
%%%%%%%%%%%%%%%%%%%%%%%%%%%%%%%%%%%%%%%%%%%%%%%
%%%%%%%%%%%%%%%%%%%%%%%%%%%%%%%%%%%%%%%%%%%%%%%

\section{Superpotentials for the other classical gauge groups}

The natural question now is whether our results can be generalised
to other known cases of dynamically generated superpotentials
for ${\cal N}=1$ super Yang--Mills theories.
Indeed, field theories with $USp(2N_c)$
\footnote{Our convention is such that $USp(2)\cong SU(2)$.} and $SO(N_c)$ gauge group
with $N_f$ flavours have been discussed in the literature.

These gauge groups are realised on branes lying right on top of the
orientifold plane. The first question to address is how the
instanton zero mode structure is modified. So imagine starting with a stack of
$N_c$ branes and an instanton wrapping the same three-cycle on the
internal manifold. As outlined in chapter~\ref{seca} there are $N_c$
fermionic zero modes $\beta_c$ and 2$N_c$ bosonic zero modes $b_c$, $\bar{\widetilde{b}}_c$
from strings starting on the instanton and ending on the D6-brane as well as $N_c$
fermionic zero modes $\widetilde{\beta}_c$ and 2$N_c$ bosonic zero modes $\bar{b}_c$, $\widetilde{b}_c$
from strings starting on the D6-brane and ending on the instanton. Upon adding an
orientifold plane wrapping the same cycle as the branes (and the instanton) only
one linear combination of $\beta_c$ and $\widetilde{\beta}_c$, $b_c$ and $\widetilde{b}_c$ as well
as $\bar{b}_c$ and $\bar{\widetilde{b}}_c$ is kept. This effectively amounts to
identifying $\beta_c$ with $-\widetilde{\beta}_c$, $b_c$ with $-\widetilde{b}_c$
and $\bar{b}_c$ with $-\bar{\widetilde{b}}_c$.

%%%%%%%%%%%%%%%%%%%%%%%%%%%%%%%%%%%%%%%%%%%%%%%
%%%%%%%%%%%%%%%%%%%%%%%%%%%%%%%%%%%%%%%%%%%%%%%
%%%%%%%%%%%%%%%%%%%%%%%%%%%%%%%%%%%%%%%%%%%%%%%
%%%%%%%%%%%%%%%%%%%%%%%%%%%%%%%%%%%%%%%%%%%%%%%

\subsection{Symplectic gauge group}

Consider now a setup where the orientifold is such that the gauge group
on the 2$N_c$ branes lying on top of it is $USp(2N_c)$. Due to the
Dirichlet boundary conditions for the strings attached to the E$2$
in the four non-compact dimensions, the gauge symmetry on the instanton is
$SO(1)$.\footnote{Note that the auxiliary group for field theory instantons in symplectic
(and orthogonal, see next section) gauge groups \cite{Hollowood:1999ev} is nicely reproduced in string theory.}
The gauge symmetry on E$2$ being trivial, we do not expect D-term 
and F-term constraints for E$2$. And indeed, as can be found for instance in 
\cite{Hollowood:1999ev}, there are neither bosonic nor fermionic ADHM constraints.
The stringy interpretation of this is the following: At least one field
in each interaction term that realises the ADHM constraints in \cite{Billo:2002hm} is odd under the
orientifold projection (see also \cite{Argurio:2007vq}). This is true
in particular for the instanton zero modes $\ov\theta_i$ and thus the corresponding ADHM constraints are absent.
As explained above there are now 2$N_c$ fermionic zero modes $\beta_c$
and 4$N_c$ bosonic zero modes $b_c$, $\bar{b}_c$. All zero modes $\beta_c$, $b_c$ and $\bar{b}_c$
transform in the fundamental representation of $USp(2N_c)$.

\begin{figure}[ht]
\begin{center}

\setlength{\unitlength}{0.75pt}
\begin{picture}(270,200)(-135,-25)

\thicklines\put(-100,0){\circle{20}}
\thicklines\put(0,150){\circle*{20}}
\thicklines\put(92,-8){\framebox(16,16){}}

\thicklines\put(-80,-5){\line(1,0){160}}
\thicklines\put(-85,15){\vector(2,3){78}}
\thicklines\put(-14,137){\vector(-2,-3){78}}
\thicklines\put(7,132){\vector(2,-3){78}}
\thicklines\put(92,20){\vector(-2,3){78}}

\thicklines\multiput(-138,0)(10,0){3}{\line(1,0){5}}
\thicklines\multiput(-87,0)(10,0){18}{\line(1,0){5}}
\thicklines\multiput(111,0)(10,0){3}{\line(1,0){5}}

\put(-135,8){$N_c$}
\put(115,8){$E2$}
\put(20,148){$N_f$}

\put(-50,-24){$\beta_c\qquad b_c\qquad\overline{b}_c$}

\put(-78,85){$\widetilde{\Phi}_{fc}$}
\put(-40,65){$\Phi_{cf}$}

\put(26,65){$\widetilde{\lambda}_f$}
\put(58,85){$\lambda_f$}

\end{picture}

\caption{Brane configuration for an instanton in an $USp(2N_c)$ gauge theory. The dashed line 
represents the orientifold plane.} \label{fig_02_so}
\end{center}
\end{figure}
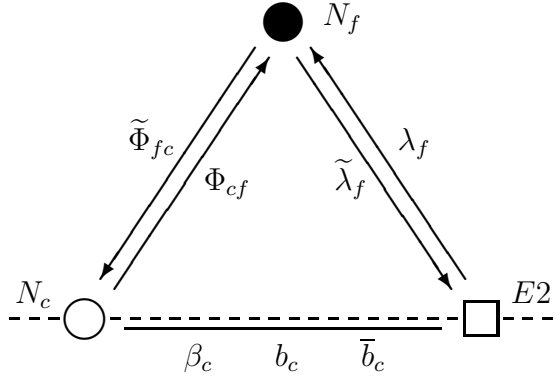

Furthermore, we denote the invariant matter strings from the 
colour brane to the flavour brane by $\Phi$, those from the 
flavour to the colour brane by $\widetilde{\Phi}$ and the fermionic zero modes
from strings between the flavour brane and the instanton by
$\lambda_f$ and $\widetilde{\lambda}_f$. All these conventions are shown in 
the extended quiver diagram in figure \ref{fig_02_so}.

In analogy to the $U(1)$ instanton case, to 
lowest order in $\alpha'$ the possible disc amplitudes with insertion of fermionic zero modes are
\eq{
  &\beta_c \,\ov{\widetilde{\Phi}}_{cf} \,\widetilde{\lambda}_f
  +\lambda_f\,\ov{\Phi}_{fc}\,\widetilde{\beta}_c 
  =\beta_c \,\ov{\widetilde{\Phi}}_{cf}\, \widetilde{\lambda}_f
  +\lambda_f\,\ov{\Phi}_{fc}\left(-\beta\right)_c 
  =\beta_c\left[\, \ov{\widetilde{\Phi}}\,,\,\ov{\Phi}^T\, \right]_{cf} 
  \left[ \begin{array}{c} \widetilde{\lambda} \\ \lambda 
  \end{array}\right]_f 
}
where we again omitted the VEV brackets for the matter fields. Note that $[\, \ov{\widetilde{\Phi}}\,,\,\ov{\Phi}^T\,]_{cf}$ is a $2N_c\times 2N_f$ matrix since the fields $\ov{\widetilde{\Phi}}_{cf}$ and $\ov{\Phi}_{fc}$ both transform in bifundamental representations of $USp(2N_c)\times U(N_f)$. Since the ADHM constraints are absent, the integration over fermionic zero modes can easily be performed 
\eq{
  \label{sp_ferm}
  &\int \prod_{c=1}^{2 N_c} d\beta_c 
  \prod_{f=1}^{N_f} d\lambda_f d\widetilde{\lambda}_f\;
  \exp\left(\beta_c\left[\, \ov{\widetilde{\Phi}}\,,\,\ov{\Phi}^T\, 
  \right]_{cf} \left[
  \begin{array}{c} \widetilde{\lambda} \\ \lambda \end{array}\right]_f
  \right) 
  = \det \left[\, \ov{\widetilde{\Phi}}\,,\,\ov{\Phi}^T\, \right] \; .
}  
However, in order to find a non-vanishing result the following constraint has to be imposed:
\eq{
  N_c=N_f \;.
}
Furthermore, we have to impose the D-term constraint for the matter fields. This implies for the VEVs that
\eq{
  \label{d_term_sp}
  \Phi_{cf} = \ov{\Phi}^T_{cf}
  \qquad\mbox{and}\qquad
  \widetilde{\Phi}_{fc} = \ov{\widetilde{\Phi}}^T_{fc} \;.
}

\bigskip
Let us now consider the bosonic zero mode integration. To lowest order in $\alpha'$, the contributing disc amplitudes with insertion of bosonic zero modes lead to
\eq{
  &\textstyle{\frac{1}{2}}\;
  b_c\left( \Phi_{cf}\, 
  \ov{\Phi}_{fc'} + \widetilde{\ov{\Phi}}_{cf}\, 
  \widetilde{\Phi}_{fc'} \right) \ov b_{c'} 
  + \textstyle{\frac{1}{2}}\;\ov{\widetilde b}_c \left( \Phi_{cf}\, 
  \ov{\Phi}_{fc'} 
  + \widetilde{\ov{\Phi}}_{cf}\, \widetilde{\Phi}_{fc'} \right) 
  \widetilde b_{c'} \\
  =&\textstyle{\frac{1}{2}}\;
  b_c\left( \Phi_{cf}\, 
  \ov{\Phi}_{fc'} + \widetilde{\ov{\Phi}}_{cf}\, 
  \widetilde{\Phi}_{fc'} \right) \ov b_{c'} 
  + \textstyle{\frac{1}{2}}\;\bigl( -\ov{b}_c \bigr) \left( \Phi_{cf}\, 
  \ov{\Phi}_{fc'} 
  + \widetilde{\ov{\Phi}}_{cf}\, \widetilde{\Phi}_{fc'} \right) 
  \bigl( -b_c \bigr)\\
  =&\textstyle{\frac{1}{2}}\;
  b_c\,\Bigl( \Phi_{cf}\, 
  \ov{\Phi}_{fc'} + \widetilde{\ov{\Phi}}_{cf}\, 
  \widetilde{\Phi}_{fc'} 
  +\left(\Phi_{cf}\, \ov{\Phi}_{fc'}\right)^T
  +\left(\widetilde{\ov{\Phi}}_{cf}\,\widetilde{\Phi}_{fc'}\right)^T
  \,\Bigr) \,\ov b_{c'} \\ 
  =& b \left[ \Phi\Phi^T + \widetilde{\Phi}^T\widetilde{\Phi}
  \right] \ov{b} \;,
} 
where in the last line the constraint \eqref{d_term_sp} was employed and we again omitted the VEV brackets. Note that $\Phi\Phi^T$ and $\widetilde{\Phi}^T\widetilde{\Phi}$ are $2N_c\times2N_c$ matrices. Since there are no bosonic ADHM constraints, the integral over bosonic zero modes to be performed is
\eq{
  &\int \prod_{c=1}^{N_c} db_c\: d\ov{b}_c\:
  \exp\left( b \left[ \Phi\Phi^T + \widetilde{\Phi}^T\widetilde{\Phi}
  \right] \ov{b}
  \right)  
  =\frac{1}{\det \left[\Phi\Phi^T + \widetilde{\Phi}^T
  \widetilde{\Phi} \right]} \;.
}
Combining this result with the one from the fermionic integration \eqref{sp_ferm} together with \eqref{d_term_sp}, one obtains
\eq{
  \frac{ \det \left[ \,\widetilde{\Phi}^T,\Phi \,\right] }
  {\det \left[\Phi\Phi^T + \widetilde{\Phi}^T\widetilde{\Phi}\right]}
  =\frac{ \det \left[ \,\widetilde{\Phi}^T,\Phi \,\right] }
  {\det \left[\Bigl[ \,\widetilde{\Phi}^T,\Phi \, \Bigr] 
  \Bigl[ \begin{array}{cc}  0&\eins_{N_c}\\ 
  \eins_{N_c}&0\end{array}\Bigr]
  \Bigl[ \,\widetilde{\Phi}^T,\Phi \, \Bigr]^T \right]}
  =-\frac{1}{\det \left[ \,\widetilde{\Phi}^T,\Phi \, \right]} \;.
}  
Note however, if we define $Q=[ \,\widetilde{\Phi}^T,\Phi \, ]$ we 
can bring this result into the following form
\eq{
  -\frac{1}{\det Q}
  =-\frac{1}{\sqrt{ \det\left[ Q 
  \Bigl[\begin{array}{cc}0 & \eins_{N_c} \\ -\eins_{N_c} 
  & 0\end{array}\Bigr]Q^T\right]}}
  =-\frac{1}{\mbox{Pfaff}\left[ \,Q\, J\, Q^T\, \right]} \; ,
}
where $J = \Bigl[\begin{array}{cc}0 & \eins_{N_c} \\ -\eins_{N_c} & 0\end{array}\Bigr]$.
Rescaling the field $Q$ such that it is canonically normalised, collecting all
dimensionful parameters and absorbing all numerical factors into $\Lambda$
the superpotential becomes
\eq{
  S_W=\int d^4 x\, d^2\theta \,\,   
  {\Lambda^{2N_f+3} \over  {\mbox{Pfaff~}\left( Q J Q^T \right)} } \;. 
}
This is precisely the form of the superpotential found by Intriligator and Pouliot in their work \cite{Intriligator:1995ne}.

%%%%%%%%%%%%%%%%%%%%%%%%%%%%%%%%%%%%%%%%%%%%%%%
%%%%%%%%%%%%%%%%%%%%%%%%%%%%%%%%%%%%%%%%%%%%%%%
%%%%%%%%%%%%%%%%%%%%%%%%%%%%%%%%%%%%%%%%%%%%%%%
%%%%%%%%%%%%%%%%%%%%%%%%%%%%%%%%%%%%%%%%%%%%%%%

\subsection{Orthogonal gauge group}

If the orientifold plane has the opposite charge than in the previous case,
the gauge group on a stack of $N_c$ D6-branes lying on top of the O6 plane is
$SO(N_c)$ and the gauge group on the instanton wrapping the same cycle is
$USp(2)$ (In this case, the smallest invariant instanton configuration is given
by two E2-branes on top of each other). The flavour brane stack is realised by $N_f$ D6-branes
wrapping a three-cycle which is distinct from the one the colour branes wrap, but also
invariant under the orientifold projection. The gauge group is assumed to be $SO(N_f)$.
There is thus one chiral supermultiplet $\Phi$ transforming in the vector representations of
both $SO(N_c)$ and $SO(N_f)$.

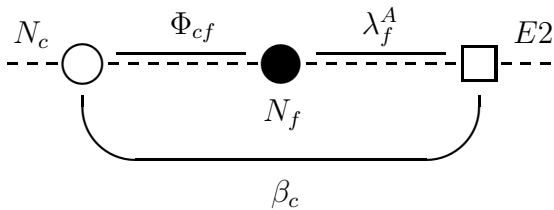
\begin{figure}[ht]
\begin{center}

\setlength{\unitlength}{0.75pt}
\begin{picture}(270,120)(-135,-80)

\thicklines\put(-100,0){\circle{20}}
\thicklines\put(0,0){\circle*{20}}
\thicklines\put(92,-8){\framebox(16,16){}}

\thicklines\put(-83,5){\line(1,0){65}}
\thicklines\put(18,5){\line(1,0){65}}

\thicklines\multiput(-138,0)(10,0){3}{\line(1,0){5}}
\thicklines\multiput(-87,0)(10,0){18}{\line(1,0){5}}
\thicklines\multiput(111,0)(10,0){3}{\line(1,0){5}}

\put(-135,10){$N_c$}
\put(117,10){$E2$}
\put(-9,-29){$N_f$}

\put(-56,15){$\Phi_{cf} \hspace{55pt} \lambda^A_f$}
\put(-5,-70){$\beta_c$}

\thicklines\put(0,-16){\oval(200,65)[b]}

\end{picture}

\caption{Brane configuration for an instanton in an $SO(N)$ 
gauge theory. The dashed line represents the orientifold 
plane.} \label{fig_03_sp}
\end{center}
\end{figure}

In the instanton-colour brane sector there are 2$N_c$ fermionic zero modes $\beta_c^A$
and 4$N_c$ bosonic zero modes $b_c^A$, $\bar{b}_c^A$, where $c=1,\ldots,N_c$ is the $O(N_c)$ index
and $A=1,\ldots,2$ the $USp(2)$ one. In addition there are 2$N_f$ fermionic zero modes $\lambda_f^A$
($f=1,\ldots,N_f$) from strings stretching between the instanton and the flavour brane.
The brane configuration together with all massless modes is shown in the extended
quiver diagram in  figure \ref{fig_03_sp}.

The D- and F-terms are adjoint valued such that in this case there are $3 \times 3 = 9$ bosonic
ADHM constraints in agreement with the field theory instanton construction \cite{Hollowood:1999ev}.
Analogously, the fermionic ADHM constraints are adjoint-valued, so there are $2 \times 3 = 6$ of them. The fermionic zero mode integral to be evaluated is thus
\eq{
  &\int \prod_{A=1}^{2}\, \prod_{c=1}^{N_c} d\beta_c^A 
  \prod_{f=1}^{N_f} 
  d\lambda_f^A \;
   \delta^{(3)}(b_c^A \sigma^i_{AB} \beta_c^B)\,
   \delta^{(3)}(\bar{b}_c^A \sigma^i_{AB} \beta_c^B)\,
   \exp \left( \beta_c^A \Phi_{cf} \lambda_f^B \right). 
}
Here, the $\sigma^i$ are the three Pauli matrices. Upon evaluating this integral,
one finds the condition $N_f = N_c - 3$ for the generation of a superpotential from
instantons. This is in agreement with the field theory result \cite{Intriligator:1995id}.

The explicit evaluation of the various bosonic and fermionic zero mode integrals, including the ADHM constraints, is non-trivial and we leave this for future work. 
Eventually one should recover the known field
theory result for the superpotential  \cite{Intriligator:1995id}
\eq{
S_W=\int d^4 x\, d^2\theta \,\,   
        {\Lambda^{2N_f+3} \over {\det} [\Phi_{fc}^T \Phi_{cf'}]  } \;. 
}

%%%%%%%%%%%%%%%%%%%%%%%%%%%%%%%%%%%%%%%%%%%%%%%
%%%%%%%%%%%%%%%%%%%%%%%%%%%%%%%%%%%%%%%%%%%%%%%
%%%%%%%%%%%%%%%%%%%%%%%%%%%%%%%%%%%%%%%%%%%%%%%
%%%%%%%%%%%%%%%%%%%%%%%%%%%%%%%%%%%%%%%%%%%%%%%
%%%%%%%%%%%%%%%%%%%%%%%%%%%%%%%%%%%%%%%%%%%%%%%
%%%%%%%%%%%%%%%%%%%%%%%%%%%%%%%%%%%%%%%%%%%%%%%
%%%%%%%%%%%%%%%%%%%%%%%%%%%%%%%%%%%%%%%%%%%%%%%
%%%%%%%%%%%%%%%%%%%%%%%%%%%%%%%%%%%%%%%%%%%%%%%

\section{Conclusions}

In this paper, for a local intersecting
D6-brane configuration giving rise to SQCD with $N_f=N_c-1$
flavours,   we have explicitly computed the one E2-instanton
contribution to the superpotential, where the E2-brane
lies on top of the stack of $N_c$ colour branes so that
in the field theory limit it can be interpreted
as a gauge instanton.
In the field theory limit, we indeed recovered the known ADS
superpotential, supporting the point of view that
string theory knows about non-perturbative
field theory effects. 
However, when one simply adds such field theory terms
to string or supergravity based superpotentials
one must keep in mind their limitations.
Moving away from the field theory limit, there
will appear stringy corrections, which one must control.
These are for instance contributions from massive states in the
one-loop Pfaffians. 
Moreover from the string perspective the gauge instanton,
i.e. E2 on top of D$6_c$, is only one of the many
possible E2-instanton contributions to the superpotential.
There can be  many other rigid special Lagrangian three-cycles
in a Calabi--Yau manifold, which can support E2-instantons. These will also give rise
to moduli dependent superpotentials, which cannot be
seen in pure field theory. In an honest computation
one must either take all these effects into account
or must control them in certain limits of the moduli space.

We have carried out the actual instanton computation in quite some
detail, firstly to show that such string space-time instanton
computations can indeed be performed explicitly and secondly 
to provide the mathematical means to actually do it.
We think that our methods are applicable to more general
instanton configurations, either admitting  a field theory limit
or being completely stringy. It would be interesting
to determine ADS like superpotentials for supersymmetric
gauge theories with more general matter fields, containing
for instance also symmetric and anti-symmetric matter.
It would also be interesting 
to derive explicitly the superpotential
induced by  gaugino condensation from string theory,
i.e. for $N_f<N_c-1$ in the case of SQCD with gauge group $SU(N_c)$.

%%%%%%%%%%%%%%%%%%%%%%%%%%%%%%%%%%%%%%%%%%%%%%%
%%%%%%%%%%%%%%%%%%%%%%%%%%%%%%%%%%%%%%%%%%%%%%%
%%%%%%%%%%%%%%%%%%%%%%%%%%%%%%%%%%%%%%%%%%%%%%%
%%%%%%%%%%%%%%%%%%%%%%%%%%%%%%%%%%%%%%%%%%%%%%%
%%%%%%%%%%%%%%%%%%%%%%%%%%%%%%%%%%%%%%%%%%%%%%%
%%%%%%%%%%%%%%%%%%%%%%%%%%%%%%%%%%%%%%%%%%%%%%%
%%%%%%%%%%%%%%%%%%%%%%%%%%%%%%%%%%%%%%%%%%%%%%%
%%%%%%%%%%%%%%%%%%%%%%%%%%%%%%%%%%%%%%%%%%%%%%%

\newpage
{\noindent  {\Large \bf Acknowledgements}}
\vskip 0.5cm
We would like to thank Timo Weigand for interesting discussions and Angel Uranga and Luis Ib\'{a}\~{n}ez for helpful correspondence on the $\ov \theta_i$ zero modes. We furthermore thank the referee for questions and clarifying comments.
This work is supported in part by the European Community's Human
Potential Programme under contract MRTN-CT-2004-005104 `Constituents,
fundamental forces and symmetries of the universe'.

%%%%%%%%%%%%%%%%%%%%%%%%%%%%%%%%%%%%%%%%%%%%%%%
%%%%%%%%%%%%%%%%%%%%%%%%%%%%%%%%%%%%%%%%%%%%%%%
%%%%%%%%%%%%%%%%%%%%%%%%%%%%%%%%%%%%%%%%%%%%%%%
%%%%%%%%%%%%%%%%%%%%%%%%%%%%%%%%%%%%%%%%%%%%%%%
%%%%%%%%%%%%%%%%%%%%%%%%%%%%%%%%%%%%%%%%%%%%%%%
%%%%%%%%%%%%%%%%%%%%%%%%%%%%%%%%%%%%%%%%%%%%%%%
%%%%%%%%%%%%%%%%%%%%%%%%%%%%%%%%%%%%%%%%%%%%%%%
%%%%%%%%%%%%%%%%%%%%%%%%%%%%%%%%%%%%%%%%%%%%%%%

\clearpage
\begin{appendix}

%%%%%%%%%%%%%%%%%%%%%%%%%%%%%%%%%%%%%%%%%%%%%%%
%%%%%%%%%%%%%%%%%%%%%%%%%%%%%%%%%%%%%%%%%%%%%%%
%%%%%%%%%%%%%%%%%%%%%%%%%%%%%%%%%%%%%%%%%%%%%%%
%%%%%%%%%%%%%%%%%%%%%%%%%%%%%%%%%%%%%%%%%%%%%%%
%%%%%%%%%%%%%%%%%%%%%%%%%%%%%%%%%%%%%%%%%%%%%%%
%%%%%%%%%%%%%%%%%%%%%%%%%%%%%%%%%%%%%%%%%%%%%%%
%%%%%%%%%%%%%%%%%%%%%%%%%%%%%%%%%%%%%%%%%%%%%%%
%%%%%%%%%%%%%%%%%%%%%%%%%%%%%%%%%%%%%%%%%%%%%%%

\section{Some relations in linear algebra}

%%%%%%%%%%%%%%%%%%%%%%%%%%%%%%%%%%%%%%%%%%%%%%%
%%%%%%%%%%%%%%%%%%%%%%%%%%%%%%%%%%%%%%%%%%%%%%%
%%%%%%%%%%%%%%%%%%%%%%%%%%%%%%%%%%%%%%%%%%%%%%%
%%%%%%%%%%%%%%%%%%%%%%%%%%%%%%%%%%%%%%%%%%%%%%%

\subsection{General formulae}
\label{app_gen_form}

For the reader's benefit, in this subsection we summarise some well-known formulae which we have used in various places in the main text.
\begin{itemize}

\item The derivative of the determinant of a matrix $A$ with respect to the matrix element $A_{i,j}$ is (Jacobi)
\eq{
  \label{app_det_deriv}
  \frac{\partial}{\partial A_{i,j}} \det A 
  = \det A \cdot \left(A^{-1}\right)_{j,i} \;. 
}

\item The determinant of a block-matrix with mutually commuting matrices $A,B,C,D$ can be computed using the following formula (Strassen)
\begin{multline}
  \label{app_det_inv}
  \left[ \begin{array}{cc} A & B \\ C & D \end{array} \right]^{-1} \\
  =\left[ \begin{array}{cc} 
  A^{-1} + A^{-1}B\left(D-CA^{-1}B\right)^{-1}CA^{-1} &
  -A^{-1}B\left(D-CA^{-1}B\right)^{-1} \\ 
  -\left(D-CA^{-1}B\right)^{-1}CA^{-1} &
  \left(D-CA^{-1}B\right)^{-1}
  \end{array} \right] \;.
\end{multline}

\item If the matrices $A,B,C,D$ mutually commute the following holds
\eq{
  \label{app_det_block}
  \det \left[ \begin{array}{cc} A & B \\ C & D \end{array} \right]
  = \det\left( AD-BC \right) \;.
}

\item For an $N\times N$ matrix the following relation holds, where, as in the main text, $A|_{k,i}$ denotes the matrix $A$ without row $k$ and column $i$
\eq{
  \label{app_det_sum}
  \det A \; \delta_{i,j} = \sum_{k=1}^N \left(-1\right)^{i+k}
   A_{k,j}\; \det A\lvert_{k,i} \;.
}

\item For an $N\times (N-1)$ matrix $A$ and an $(N-1)\times N$ matrix $B$ it is clear that the determinant of the $N\times N$ matrix $AB$ vanishes:
\eq{
  \label{app_det_rank}
  \det AB=0 \;.
}

\item The $(i,j)$th element of the inverse matrix $A^{-1}$ is
\eq{
  \label{app_det_inv02}
  A^{-1}_{i,j} = \left(-1\right)^{i+j} \frac{\det A\lvert_{j,i}}{\det A}
  \;.
}

\item For $N\times N$ matrices $A$ and $B$ the following relation holds
\eq{
  \label{app_det_sum02}
  \sum_k \det A|_{i,k}\cdot\det B|_{k,j} = \det AB|_{i,j} \;.
}

\end{itemize}

%%%%%%%%%%%%%%%%%%%%%%%%%%%%%%%%%%%%%%%%%%%%%%%
%%%%%%%%%%%%%%%%%%%%%%%%%%%%%%%%%%%%%%%%%%%%%%%
%%%%%%%%%%%%%%%%%%%%%%%%%%%%%%%%%%%%%%%%%%%%%%%
%%%%%%%%%%%%%%%%%%%%%%%%%%%%%%%%%%%%%%%%%%%%%%%

\subsection{A determinant formula}
\label{app_detform}

Let $A$ be an $N \times (N-1)$ matrix and $B$ be an $(N-1) \times N$ matrix. Then define
\eq{
  \widetilde{A} = \left[ A_{N,(N-1)}, 0_{N,1} \right] 
  \qquad \mbox{and} \qquad
  \widetilde{B} = \left[ \begin{array}{c} B_{(N-1),N} \\ 0_{1,N} 
  \end{array} \right] 
}
and note that $AB = \widetilde{A} \widetilde{B}$. Then one finds using \eqref{app_det_sum02}
\begin{align}
  \label{app_detform02}
  &\sum_{i=1}^N \det \Bigl( AB\bigl( \bigl(AB+\alpha\,\eins\bigr)^2 
  +\beta^2\,\eins
  \bigr) \Bigr)\Bigr\rvert_{i,i} \\
  =&\sum_{i=1}^N \det \Bigl( AB\bigl( (AB)^2 +2\alpha AB + \alpha^2\eins
  +\beta^2\eins \bigr) \Bigr)\Bigr\rvert_{i,i} \nonumber \\
  =&\sum_{i,j=1}^N \det AB\bigr\rvert_{i,j} \;
  \det \bigl( (AB)^2 +2\alpha AB + \alpha^2\,\eins +\beta^2\,\eins 
  \bigr)\bigr\rvert_{j,i}
  \nonumber\\
  =&\sum_{i,j=1}^N \det \widetilde{A}\widetilde{B}\bigr\rvert_{i,j} \;
  \det \bigl( (\widetilde{A}\widetilde{B})^2 +2\alpha 
  \widetilde{A}\widetilde{B} 
  + \alpha^2\,\eins +\beta^2\,\eins \bigr)\bigr\rvert_{j,i} \nonumber\\
  =&\sum_{i,j,k=1}^N \det \widetilde{A}\bigr\rvert_{i,k}\; 
  \det\widetilde{B}\bigr\rvert_{k,j} \;
  \det \bigl( (\widetilde{A}\widetilde{B})^2 +2\alpha 
  \widetilde{A}\widetilde{B} 
  + \alpha^2\,\eins +\beta^2\,\eins \bigr)\bigr\rvert_{j,i} \;. 
  \nonumber
\end{align}
Now note that $\det \widetilde{A}\bigr\rvert_{i,k}$ and $\det\widetilde{B}\bigr\rvert_{k,j}$ are only nonzero for $k=N$. Therefore one can write
\begin{align}
  &\sum_{i,j=1}^N  
  \det\widetilde{B}\bigr\rvert_{N,j} \;
  \det \bigl( (\widetilde{A}\widetilde{B})^2 +2\alpha 
  \widetilde{A}\widetilde{B} 
  + \alpha^2\eins +\beta^2\eins \bigr)\bigr\rvert_{j,i}\;
  \det \widetilde{A}\bigr\rvert_{i,N} \nonumber \\ 
  =&\det \Bigl( \widetilde{B}
  \bigl( (\widetilde{A}\widetilde{B})^2 +2\alpha 
  \widetilde{A}\widetilde{B} 
  + \alpha^2\eins +\beta^2\eins 
  \bigr)\widetilde{A}\Bigr)\Bigr\rvert_{N,N} 
  \nonumber \\
  \label{app_detform01}
  =&\det\Bigl(
  (\widetilde{B}\widetilde{A})^3 +2\alpha 
  (\widetilde{B}\widetilde{A})^2 
  + \left(\alpha^2 +\beta^2\right)\widetilde{B}\widetilde{A}
  \Bigr)\Bigr\rvert_{N,N} \;.
\end{align}    
Finally, observe that
\eq{
  \widetilde{B}\widetilde{A} = \left[
   \begin{array}{cc} BA & 0 \\ 0 & 0 \end{array} 
   \right] \nonumber
}
which then implies that \eqref{app_detform01} is equal to 
\begin{align}
  &\det\Bigl(
  (BA)^3 +2\alpha (BA)^2 + \left(\alpha^2 +\beta^2\right)BA  \Bigr) 
  \nonumber\\
  \label{app_detform03}
  =&\det\Bigl( BA \bigl( \bigl( BA +\alpha \eins\bigr)^2 + \beta^2 
  \eins\bigr)
  \Bigr) \;. 
\end{align}
This computation shows that \eqref{app_detform02} equal to \eqref{app_detform03}.

%%%%%%%%%%%%%%%%%%%%%%%%%%%%%%%%%%%%%%%%%%%%%%%
%%%%%%%%%%%%%%%%%%%%%%%%%%%%%%%%%%%%%%%%%%%%%%%
%%%%%%%%%%%%%%%%%%%%%%%%%%%%%%%%%%%%%%%%%%%%%%%
%%%%%%%%%%%%%%%%%%%%%%%%%%%%%%%%%%%%%%%%%%%%%%%

\subsection{A formula concerning characteristic polynomials}
\label{app_charpol}

Let $A$ be an $N \times (N-k)$ matrix and $B$ be an $(N-k) \times N$ matrix for $k=1,\ldots, (N-1)$. Then define
\eq{
  \widetilde{A} = \left[ A_{N,(N-k)}, 0_{N,k} \right] 
  \qquad \mbox{and} \qquad
  \widetilde{B} = \left[ \begin{array}{c} B_{(N-k),N} \\ 0_{k,N} 
  \end{array} \right] 
}
and note that $AB=\widetilde{A}\widetilde{B}$. Furthermore, denote the characteristic polynomial of a matrix $M$ as $\chi_{M}(\sigma)$ and recall that $\chi_{MN}(\sigma)=\chi_{NM}(\sigma)$ for square matrices $M,N$. Then it is easy to see that
\eq{
  \det \bigl[ AB + \lambda\,\eins \bigr]  
  =&\det \bigl[ \widetilde{A}\widetilde{B} + \lambda\,\eins \bigr] \\ 
  =& \chi_{\widetilde{A}\widetilde{B}}\left( -\lambda \right) \\
  =& \chi_{\widetilde{B}\widetilde{A}}\left( -\lambda \right) \\
  =&\det \bigl[ \widetilde{B}\widetilde{A} + \lambda\,\eins \bigr] \\ 
  =&\det \left[ \begin{array}{cccc} BA + \lambda  \\
    & \lambda \\
    && \ddots \\
    &&&\lambda \end{array} \right] \\ 
  =&\lambda^k \det \bigl[ BA + \lambda\,\eins \bigr] \;.
}

%%%%%%%%%%%%%%%%%%%%%%%%%%%%%%%%%%%%%%%%%%%%%%%
%%%%%%%%%%%%%%%%%%%%%%%%%%%%%%%%%%%%%%%%%%%%%%%
%%%%%%%%%%%%%%%%%%%%%%%%%%%%%%%%%%%%%%%%%%%%%%%
%%%%%%%%%%%%%%%%%%%%%%%%%%%%%%%%%%%%%%%%%%%%%%%
%%%%%%%%%%%%%%%%%%%%%%%%%%%%%%%%%%%%%%%%%%%%%%%
%%%%%%%%%%%%%%%%%%%%%%%%%%%%%%%%%%%%%%%%%%%%%%%
%%%%%%%%%%%%%%%%%%%%%%%%%%%%%%%%%%%%%%%%%%%%%%%
%%%%%%%%%%%%%%%%%%%%%%%%%%%%%%%%%%%%%%%%%%%%%%%

\end{appendix}

%%%%%%%%%%%%%%%%%%%%%%%%%%%%%%%%%%%%%%%%%%%%%%%
%%%%%%%%%%%%%%%%%%%%%%%%%%%%%%%%%%%%%%%%%%%%%%%
%%%%%%%%%%%%%%%%%%%%%%%%%%%%%%%%%%%%%%%%%%%%%%%
%%%%%%%%%%%%%%%%%%%%%%%%%%%%%%%%%%%%%%%%%%%%%%%
%%%%%%%%%%%%%%%%%%%%%%%%%%%%%%%%%%%%%%%%%%%%%%%
%%%%%%%%%%%%%%%%%%%%%%%%%%%%%%%%%%%%%%%%%%%%%%%
%%%%%%%%%%%%%%%%%%%%%%%%%%%%%%%%%%%%%%%%%%%%%%%
%%%%%%%%%%%%%%%%%%%%%%%%%%%%%%%%%%%%%%%%%%%%%%%

\clearpage
\nocite{*}
\bibliography{rev}
\bibliographystyle{utphys}

%%%%%%%%%%%%%%%%%%%%%%%%%%%%%%%%%%%%%%%%%%%%%%%
%%%%%%%%%%%%%%%%%%%%%%%%%%%%%%%%%%%%%%%%%%%%%%%
%%%%%%%%%%%%%%%%%%%%%%%%%%%%%%%%%%%%%%%%%%%%%%%
%%%%%%%%%%%%%%%%%%%%%%%%%%%%%%%%%%%%%%%%%%%%%%%
%%%%%%%%%%%%%%%%%%%%%%%%%%%%%%%%%%%%%%%%%%%%%%%
%%%%%%%%%%%%%%%%%%%%%%%%%%%%%%%%%%%%%%%%%%%%%%%
%%%%%%%%%%%%%%%%%%%%%%%%%%%%%%%%%%%%%%%%%%%%%%%
%%%%%%%%%%%%%%%%%%%%%%%%%%%%%%%%%%%%%%%%%%%%%%%

\end{document}